\documentclass[manuscript,screen]{acmart}

\AtBeginDocument{%
  }

\setcopyright{acmlicensed}
\copyrightyear{2018}
\acmYear{2018}
\acmDOI{XXXXXXX.XXXXXXX}

\acmConference[Conference acronym 'XX]{Make sure to enter the correct
  conference title from your rights confirmation emai}{June 03--05,
  2018}{Woodstock, NY}
\acmISBN{978-1-4503-XXXX-X/18/06}

\usepackage{balance}
\usepackage{multirow}
\usepackage{amsthm}
\usepackage{enumitem}
\usepackage[normalem]{ulem}
\usepackage{color}
\usepackage{booktabs}
\usepackage{multirow}
\usepackage{mdframed}

\usepackage{amsmath}
\usepackage{tcolorbox}
\usepackage{tikz}
\usetikzlibrary{decorations.pathreplacing, positioning}

\usepackage{graphicx}

\usepackage[normalem]{ulem}
\useunder{\uline}{\ul}{}

\DeclareMathOperator*{\argmin}{arg\,min}

\useunder{\uline}{\ul}{}

\newcommand{\ie}{\emph{i.e., }}
\newcommand{\eg}{\emph{e.g., }}
\newcommand{\etal}{\emph{et al.}}
\newcommand{\etc}{\emph{etc.}}

\definecolor{mycolor}{HTML}{ed028c}

\begin{document}

\title{LLM4DSR: Leveraing Large Language Model for Denoising Sequential Recommendation}


\author{Bohao Wang}
\email{bohao.wang@zju.edu.cn}
\affiliation{%
  \institution{Zhejiang University}
  \city{Hangzhou}
  \country{China}
}

\author{Feng Liu}
\email{liufeng4hit@gmail.com}
\affiliation{%
  \institution{OPPO Co Ltd}
  \city{Shenzhen}
  \country{China}
}

\author{Changwang Zhang}
\email{changwangzhang@foxmail.com}
\affiliation{%
  \institution{OPPO Co Ltd}
  \city{Shenzhen}
  \country{China}
}

\author{Jiawei Chen}
\email{sleepyhunt@zju.edu.cn}
\affiliation{%
  \institution{Zhejiang University}
  \city{Hangzhou}
  \country{China}
}

\author{Yudi Wu}
\email{wuyudi@zju.edu.cn}
\affiliation{%
  \institution{Zhejiang University}
  \city{Hangzhou}
  \country{China}
}

\author{Sheng Zhou}
\email{zhousheng_zju@zju.edu.cn}
\affiliation{%
  \institution{Zhejiang University}
  \city{Hangzhou}
  \country{China}
}

\author{Xingyu Lou}
\email{louxingyu@oppo.com}
\affiliation{%
  \institution{OPPO Co Ltd}
  \city{Shenzhen}
  \country{China}
}

\author{Jun Wang}
\email{junwang.lu@gmail.com}
\affiliation{%
  \institution{OPPO Co Ltd}
  \city{Shenzhen}
  \country{China}
}

\author{Yan Feng}
\email{fengyan@zju.edu.cn}
\affiliation{%
  \institution{Zhejiang University}
  \city{Hangzhou}
  \country{China}
}

\author{Chun Chen}
\email{chenc@zju.edu.cn}
\affiliation{%
  \institution{Zhejiang University}
  \city{Hangzhou}
  \country{China}
}

\author{Can Wang}
\email{wcan@zju.edu.cn}
\affiliation{%
  \institution{Zhejiang University}
  \city{Hangzhou}
  \country{China}
}

\renewcommand{\shortauthors}{Bohao et al.}

\begin{abstract}
  Sequential Recommenders generate recommendations based on users' historical interaction sequences. However, in practice, these collected sequences are often contaminated by noisy interactions, which significantly impairs recommendation performance. Accurately identifying such noisy interactions without additional information is particularly challenging due to the absence of explicit supervisory signals indicating noise. Large Language Models (LLMs), equipped with extensive open knowledge and semantic reasoning abilities, offer a promising avenue to bridge this information gap. However, employing LLMs for denoising in sequential recommendation presents notable challenges: 1) Direct application of pretrained LLMs may not be competent for the denoising task, frequently generating nonsensical responses;  2) Even after fine-tuning, the reliability of LLM outputs remains questionable, especially given the complexity of the denoising task and the inherent hallucinatory issue of LLMs.
  
  To tackle these challenges, we propose LLM4DSR, a tailored approach for denoising sequential recommendation using LLMs. We constructed a self-supervised fine-tuning task to activate LLMs' capabilities to identify noisy items and suggest replacements. Furthermore, we developed an uncertainty estimation module that ensures only high-confidence responses are utilized for sequence corrections. Remarkably, LLM4DSR is model-agnostic, allowing corrected sequences to be flexibly applied across various recommendation models. Extensive experiments validate the superiority of LLM4DSR over existing methods.
  
\end{abstract}

\ccsdesc[500]{Information systems~Recommender systems}
\keywords{Sequential Recommendation, Denoise, Large Language Model}

\received{20 February 2007}
\received[revised]{12 March 2009}
\received[accepted]{5 June 2009}

\maketitle

\section{Introduction}
Large Language Models (LLMs) have demonstrated remarkable capabilities in content comprehension, generation, and semantic reasoning, thereby catalyzing a revolution in artificial intelligence \cite{achiam2023gpt}. Recently, LLMs have been extensively applied in the field of Recommender Systems (RS) \cite{wu2023survey}. Some researchers have directly utilized LLMs by prompting or fine-tuning them to act as specialized recommenders \cite{liao2024llara, harte2023leveraging, bao2023tallrec}. Meanwhile, others have employed LLMs as auxiliary tools to enhance traditional recommendation models \eg serving as encoders for user/item features \cite{wei2024llmrec}, supplementary knowledge bases \cite{xi2023towards}, advanced recommendation explainers \cite{tennenholtz2023demystifying}, and conversational agents \cite{gao2023chat}.

Given the powerful capabilities of LLMs, this work explores a innovative question: \textbf{Can LLMs effectively function as denoisers for sequential recommendation?} Sequential recommendation critically relies on the accuracy of users' historical interaction sequences to predict subsequent items. However, in practice, these sequences are often contaminated by noise noise due to various factors --- \eg being attracted by clickbait \cite{wang2021clicks}, influenced by item prominent positions \cite{chen2023bias}, or from accidental interactions \cite{lin2023self}. Such noise can significantly mislead recommendation models and degrade their performance.
Figure \ref{intro_noise_ratio} provides empirical evidence of this impact: with only a small proportion (e.g., 20\%) of randomly selected noisy items, we observe performance drops of 41\% and 36\% in the sequential recommendation models SASRec and LLaRA, respectively. The importance of denoising these sequences cannot be overstated.

Despite its importance, the task of denoising is inherently challenging. A primary issue is the absence of labels for noisy data, resulting in a lack of clear signals regarding the nature of the noise. Recent approaches to denoising in sequential recommendation either rely on on human-designed heuristics \cite{zhang2024ssdrec, zhang2022hierarchical, sun2021does}, which require extensive expertise and often lack precision; or are trained jointly with recommendation models guided by the recommendation objective \cite{zhang2022hierarchical, yuan2021dual, chen2022denoising, lin2023self}, which may still be heavily influenced by the noisy data, particularly since the labels of the recommendation objective (\ie the next item) might also be contaminated. \textbf{Thus, this work explores a new avenue of leveraging LLMs for recommendation denoising, which can potentially compensate for this knowledge gap.} LLMs, with their extensive open knowledge and advanced semantic reasoning abilities \cite{brown2020language}, are equipped to identify abnormal interactions from a semantic perspective. For example, if a user's viewing history primarily consists of comedy films, an occasional interaction with a horror film might be attributable to an accidental click. LLMs can detect such anomalies based on their understanding of item semantics and common-sense knowledge. Moreover, by leveraging the generative capabilities of LLMs, they can not only identify noisy items but also potentially suggest replacements that align more closely with user preferences.

\begin{figure}[t]
  \centering
  \includegraphics[width=\linewidth]{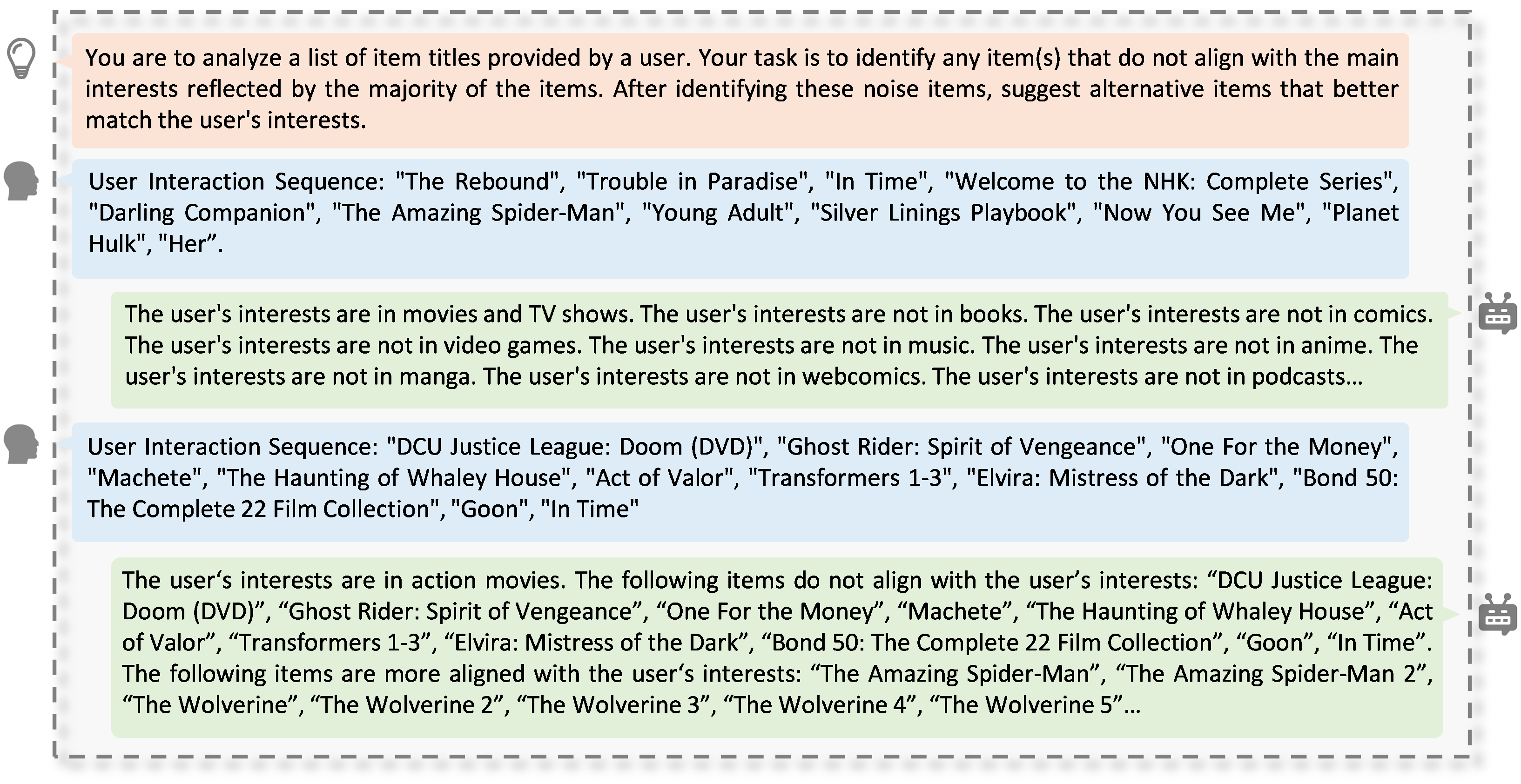}
  \caption{Failure cases of using the un-finetuned Llama3 8B model for denoising tasks. The problems in the output include meaningless output, classifying all items as noise, generating fake items, \etc 50 test data are used to generate responses with Llama 3, and the reasonableness of these responses is evaluated using GPT-4. Over 40\% of the responses were deemed nonsensical.}
  \Description{}
  \label{fail_case}
\end{figure}

However, employing pre-trained LLMs for denoising sequential recommendation also presents significant challenges:
\begin{itemize}[leftmargin=*]
  \item \textbf{Challenge 1}: \textbf{Direct application of pretrained LLMs may not be competent for denoising tasks}. 
  A common approach to utilizing LLMs involves structuring the denoising task as a prompt and directly feeding it into the LLMs. However, the significant gap between the pre-training objectives of LLMs and the specific requirements of sequence recommendation denoising can make such direct application problematic. In our preliminary experiments with the pretrained Llama3-8B model \cite{dubey2024llama}, we observed that a significant portion (over 40\%) of the model's responses were nonsensical. Figure \ref{fail_case} presents typical failure cases, where the LLM frequently misclassifies all items in a sequence as noise.
  \item \textbf{Challenge 2}: \textbf{Responses from LLMs might not always be reliable.} Even with fine-tuning, LLMs are susceptible to the notorious phenomenon of \textit{hallucination} \cite{ji2023survey}, where they might incorrectly label non-existent items as noise or suggest non-existent replacements. Moreover, given the complexity of the recommendation denoising task, the accuracy of an LLM-based denoiser still has an upper-bound. Blindly trusting LLMs for corrections may not be advisable.
\end{itemize}

To address these challenges, we propose LLM4DSR, a tailored approach for denoising sequential recommendations using LLMs. To overcome the first challenge, we adapt pre-trained LLMs to the denoising task through a self-supervised fine-tuning process. We construct an instruction dataset by replacing a proportion of items in sequences with randomly selected alternatives, which are then used for fine-tuning LLMs to identify noisy items and suggest appropriate replacements. This approach activates the potential denoising ability of LLMs. To tackle the second challenge, we introduce an uncertainty estimation module to assess the reliability of the identified noisy items, ensuring that only high-confidence judgments are used to correct sequences for subsequent recommendations. Notably, LLM4DSR is data-centric and model-agnostic, which can be considered a data preprocessing technique. The corrected sequences can be seamlessly applied across various recommendation models. Our extensive experiments demonstrate that LLM4DSR significantly enhances the performance of both traditional sequential recommendation models and advanced LLM-based models. 

The primary contributions of this work are:
\begin{itemize}
\item Introducing the novel application of LLMs in denoising sequential recommendations and identifying key challenges.
\item Proposing LLM4DSR, a novel LLM-based recommendation denoising method that utilizes self-supervised fine-tuning and an uncertainty estimation mechanism to fully exploit the capabilities of LLMs.
\item Conducting comprehensive experiments that demonstrate the superiority of LLM4DSR over state-of-the-art denoising methods across three recommendation backbone models.
\end{itemize}

The remainder of this article is organized as follows: Section \ref{Preliminary} outlines the basic definitions of the tasks and methods involved. The proposed method, LLM4DSR, is introduced in Section \ref{Method}. Section \ref{Experiments} presents our experimental results and subsequent discussions. Related work is reviewed in Section \ref{Related work}. Finally, we conclude the article and suggest future research directions in Section \ref{Conclusion}.

\begin{table*}[t]
    \centering
    \caption{Notations in this paper.}
    \begin{tabular}{ll}
    \toprule 
    \textbf{Notations}&\textbf{Descriptions}\\
    \midrule 
    $v_i$ & the $i$-th item in the item set $\mathcal V$ \\
    $S$ & the historical temporal interaction sequence \\
    $S'$ & the noiseless sequence of $S$ generated by the model \\
    $\hat{S}$ & the modified sequence of $S$ generated by randomly replacing \\
    $s_t$ & the $t$-th item in the sequence $S$ \\
    $y_t$ & whether the interaction $s_t$ is noise predicted by the model \\
    $r_t$ & the replacement item generated by the model for $s_t$ \\
    $text(s_t)$ & the textual description of the item $s_t$ \\
    $x$ & the token sequence of the input part of the prompt \\
    $z$ & the token sequence of the out part of the prompt \\
    $z_{noise}$ & the token sequence of the token sequence of the noise item given by the LLM \\
    $\Phi$ & the model parameter \\
    $T_i$ & the token sequence of the textual description of the item $s_i$ \\
    $\eta$ & the probability threshold to classify items as noise \\
    $\alpha$ & noise ratio of artificially noisy dataset \\
    $b$ & the maximum number of items modified in a single sequence \\
    $q_{ui}$ & the ranking position of the item $i$ for the user $u$. \\
    \bottomrule
    \label{notation}
    \end{tabular}
\end{table*}

\section{Preliminary}
\label{Preliminary}
In this section, we present the background of the task of sequential recommendation denoising and the large language models. Tabel \ref{notation} summarizes the notations involved in this article.

\subsection{Background of Sequential Recommendation}
Our problem definitions directly refer to recent work \cite{kang2018self, sun2019bert4rec, bao2023bi}. Given a sequential recommender system, let $\mathcal{V}=\left\{v_1, v_2, \cdots, v_{|\mathcal{V}|}\right\}$ be the set of items. We use $S^i=\left[s_1^i, \cdots, s_t^i, \cdots, s_{|S^i|}^i\right]$ to denote $i$-th historical collected interaction sequence where $s_t^i \in \mathcal{V}$ denotes the $t$-th item interacted within sequence $S^i$, and $|S^i|$ represents the sequence's length. We denote a subsequence $[s^i_j, \cdots, s^i_k]$ of $S^i$ as $S^i_{j:k}$. 
For each sequence $S^i$, sequential recommendation takes $S^i_{1:t}$ sequence data as input and outputs the item that the user is likely to interact with next, which is expected as $S^i_{t+1}$.
For the sake of brevity, we will use $S$ to replace $S^i$ in the reminder of the paper.

\begin{figure}[t]
  \centering
  \includegraphics[width=0.8\linewidth]{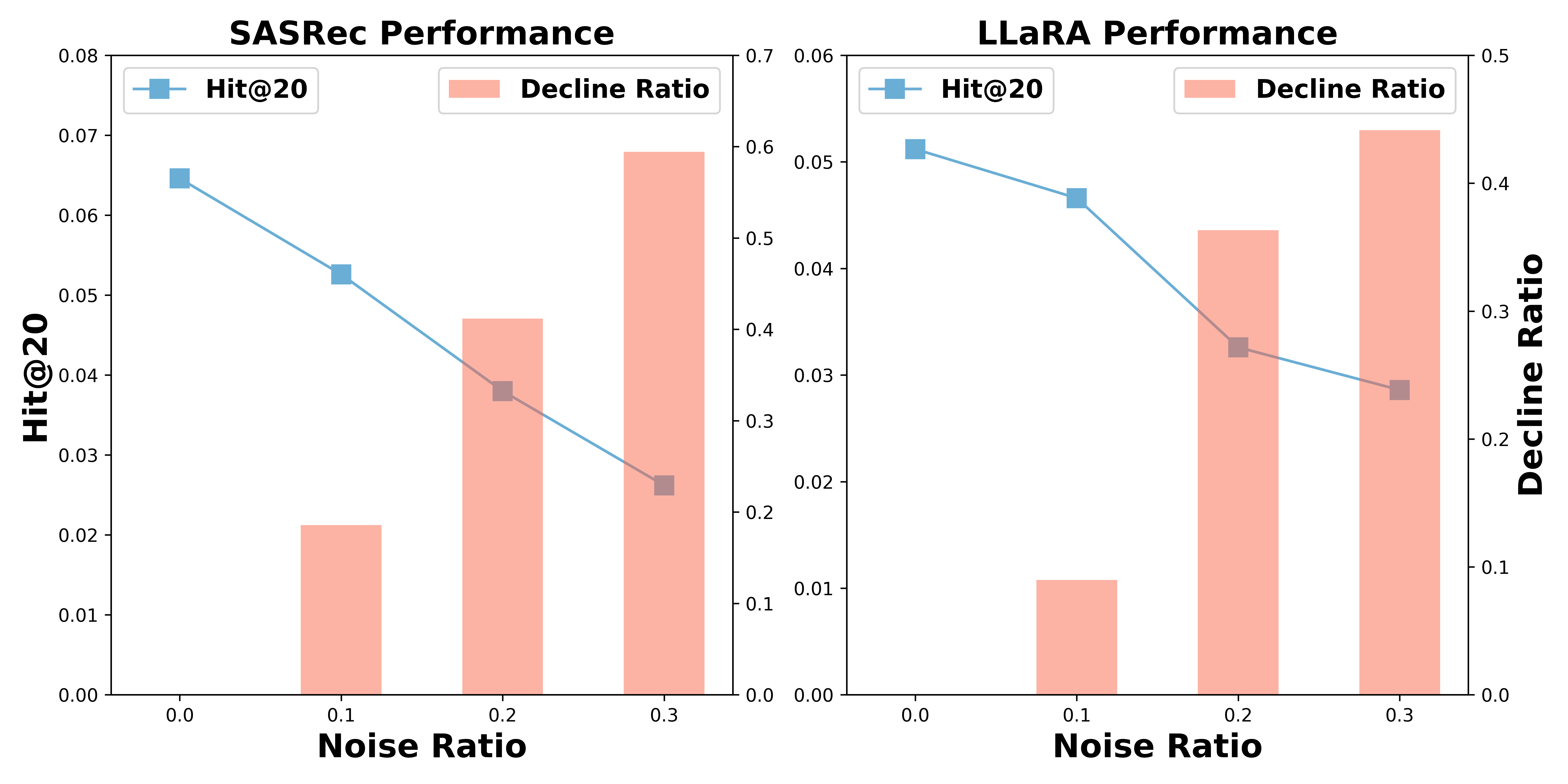}
  \caption{Performance of SASRec \cite{kang2018self} (a traditional sequential recommendation model) and LLaRA \cite{liao2024llara} (a sequential recommendation model based on LLMs) under different noise ratios and the proportion of performance degradation relative to the noise-free condition. Noise was introduced by randomly replacing each item in the sequence with a probability $\alpha$, and different levels of noise were introduced by setting different values of $\alpha$. The experiments are conducted on the Amazon Games, with dataset details provided in Section \ref{exp_dataset}. Both models exhibit significant performance degradation as the noise ratio increases.}
  \Description{}
  \label{intro_noise_ratio}
\end{figure}

\subsection{Sequential Recommendation Denoising.}
The presence of noisy interactions in user behavior sequences can impact model performance. As shown in Figure \ref{intro_noise_ratio}, both traditional sequential recommendation models and recent large language model-based recommendation models exhibit a significant decline in performance as the noise ratio increases. Therefore, developing denoising techniques for sequential recommendation is an important task.

Recent work on sequential recommendation denoising can be mainly categorized into two types: 1) the explicit denoising and 2) the implicit denoising. For a given sequence \( S \), the objective of the explicit denoising strategies \cite{sun2021does, zhang2022hierarchical, lin2023self} target at identifying and removing the noise within the sequence and ultimately generate a noiseless sequence \( S' \).
Implicit denoising methods \cite{yuan2021dual, chen2022denoising, zhou2022filter, han2024end4rec} do not directly remove items from the sequence. Instead, they reduce the impact of noise with respect to the final representation by decreasing the weight of noise items, or by filtering noise at the representation level through filters.

Traditional denoising methods encounter significant challenges due to the absence of labeled noise data. These methods often rely on heuristic prior knowledge for noise identification, which requires extensive expertise and frequently lacks precision. Alternatively, some denoising approaches employ recommendation loss as a supervisory signal during the denoising process; however, this can be adversely affected by the presence of noise. Moreover, explicit denoising methods risk information loss from sequences due to potential erroneous deletions, which can degrade model performance.

To address these limitations, this work explores leveraging LLMs in sequential recommendation denoising. The advantages of this strategy are multifaceted: 1) LLMs possess extensive open knowledge and advanced semantic reasoning abilities, which can potentially bridge the information gap in the denoising task; 2) The denoising task can be decoupled from the learning of recommendation models, serving as a data preprocessing step, thus allowing the corrected sequence to benefit various sequential recommendation models; 3) Owing to the generative ability of LLMs, the method can not only identify noisy items but also potentially suggest replacements. This can effectively reduce the impact of erroneous deletions, and the additional items introduced by the replacement operation may indeed provide useful information, benefiting sequence recommendation. We also empirically validate the effectiveness of such generation in our experiments. 

Formally, let \( S' \) denote the corrected sequence of the original sequence \( S \), where each \( s'_t \in S' \) represents the \( t \)-th item of the corrected sequence. The denoising process in this paper can be expressed as follows:
\[
s'_t = \begin{cases}
s_t, &  y_t = 0 \\
r_t, &  y_t = 1
\end{cases}
\]
where \( y_t = 1 \) indicates that the denoising model classify the item $s_t$ as noise, while \( y_t = 0 \) indicates not. \( r_t \) denotes the replacement item generated by the LLMs for $s_t$. 

\subsection{Background of Large Language Model}
Large Language Models (LLMs) have become transformative tools in natural language processing. These models, built on the transformer architecture \cite{vaswani2017attention}, effectively capture relationships in sequential data and can easily scaled to unprecedented sizes, reaching billions of parameters \cite{dubey2024llama, achiam2023gpt}. As the number of parameters increases, LLMs exhibit emergent abilities not present in traditional NLP models, leading to significant improvements in language understanding and generation tasks \cite{wei2022emergent}. Trained on vast corpora of text from diverse sources, LLMs are able to capture a wide range of linguistic patterns and contextual nuances. They have demonstrated remarkable capabilities in tasks such as translation \cite{zhu2023multilingual}, summarization \cite{basyal2023text}, and question-answering \cite{huo2023retrieving}, making them a pivotal component of modern AI research and applications.

In addition, LLMs pre-trained on extensive datasets exhibit strong zero-shot or few-shot capabilities for downstream tasks, requiring only minimal fine-tuning to quickly adapt to specific tasks \cite{zhang2023instruction}. Instruction tuning \cite{chung2024scaling} is a fine-tuning method that enables LLMs to better understand and follow human instructions. By training the model on a labeled dataset of instructional prompts and corresponding outputs, instruction tuning significantly improves the model's performance in specified instruction formats. This approach not only refines the model's outputs but also enhances its interpretability and reliability in practical applications. Specifically, the instruction data is organized in the format $(x,z)$, where \(x\) represents the text-based instruction and \(z\) represents the corresponding response. LLMs have demonstrated a significant improvement in performance on specific tasks when trained with a limited amount of instructional data \cite{zhang2023instruction}.

\begin{figure}[t]
  \centering
  \abovecaptionskip=-0pt
  \belowcaptionskip=-5pt
  \includegraphics[width=0.6\linewidth]{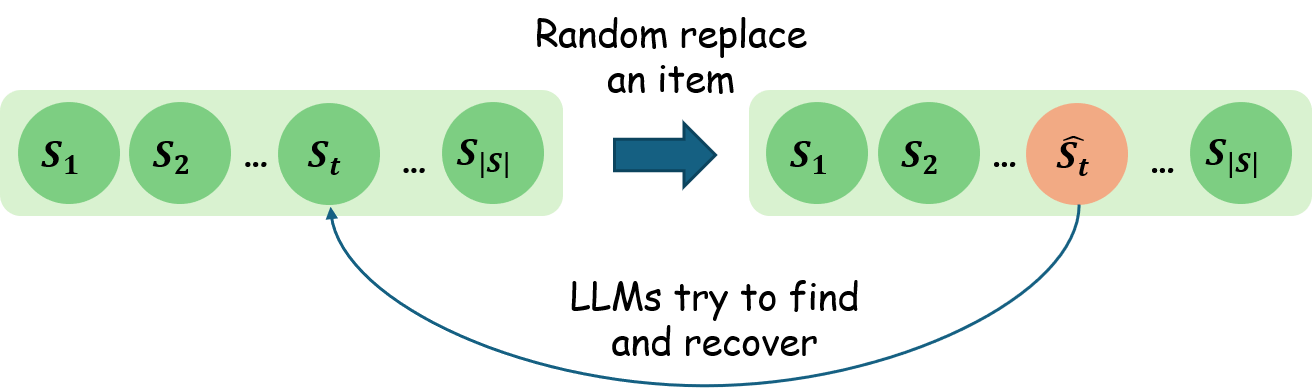}
  \caption{The self supervised task for fine-tuning LLMs as denoiser.}
  \label{self_supervise}
\end{figure}

\section{Methodology}
\label{Method}
In this section, we present the proposed LLM4DSR. We first employ a self-supervised instruction tuning approach to enable the denoising capability of the LLM (Subsection \ref{train_LLM}), and then introduce an uncertainty estimation module to improve the reliability of the denoising process (Subsection \ref{uncertainty_estimation}). Finally, we describe the procedure for replacing noisy samples with items suggested by the LLM (Subsection \ref{replace}).

\subsection{Fine-tuning LLM as a Denoiser}
\label{train_LLM}
To better activate the denoising capabilities of LLMs, we propose leveraging an instruction tuning strategy. Instruction tuning has been widely adopted in various LLM-based applications, and has proven effective in enabling LLMs to quickly adapt to new tasks \cite{zhang2023instruction}. Specifically, the template for the instruction data in this task can be formulated as follows:



\begin{tcolorbox}[colback=yellow!5!white, colframe=yellow!50!black, colbacktitle=yellow!75!black, title=\textbf{Template of Instruction Tuning}]
\textbf{Instruction}: You are to analyze a list of items provided by a user. Your task is to identify an item that do not align with the main interests reflected by the majority of the items. After identifying these noise items, suggest alternative items that better match the user's interests. \\ \\
\textbf{Input}: The user has interacted with the following items before: \colorbox{green!25}{$text(s_1), \cdots, text(\hat{s_t}), \cdots, text(s_{|S|})$} \\ \\
\textbf{Output}:\colorbox{red!25}{Noise Items:}\colorbox{green!25}{${text}(\hat{s_t})$},\colorbox{red!25}{Suggested Items:}\colorbox{green!25}{${text}(s_t)$}
\end{tcolorbox}
where $text(s_i)$ represents the textual description of item $s_i$ (e.g., titles). Here, we organize the language descriptions of sequences as instruction inputs and expect the LLMs to identify noisy items and suggest replacements. Specifically, the prompt \textbf{"The user has interacted with the following items before:"} provides the model with information about the user's historical interactions. \textbf{"Noise Items:"} guides the model to identify noise within the sequence from the semantic context. \textbf{"Suggested Items:"} encourages the model to infer the user's interests based on the sequence and suggest replacements that align more closely with the user's preferences.

However, one challenge of such an instruction fine-tuning strategy is the lack of explicit signals denoting the ground truth of the noise items and ideal replacement items. To address this, we propose leveraging a self-supervised strategy to construct instruction data. We randomly select an item $s_t$ from the original sequences and replace it with a noise item $\hat{s_t}$ that is randomly sampled from $V$. LLMs can be fine-tuned on the corrupted sequence, aiming to predict the modified noise item and make predictions of the original item. The schematic diagram of the self-supervised task is shown in Figure \ref{self_supervise}. In this way, LLMs can effectively understand the task of sequential recommendation denoising and generalize to identify potential abnormal items in the original sequence and make corrections, which would significantly boost recommendation performance.

Formally, let denote the token sequence of input part as $x$ and output part as $z$. Then we can fine-tune the LLMs on the aforementioned corpus and the objective function can be formalized as follows:
\begin{equation}
\max _{\Phi} \sum_{(x, z)} \sum_{i=1}^{|z|} \log \left(P_{\Phi}\left(z_i \mid x, z_{<i}\right)\right),
\end{equation} 
where $\Phi$ represents the model parameter and $|z|$ is the total length of token sequence of the output part. $P_{\Phi}\left(z_i \mid x, z_{<i}\right)$ signifies the probability of the token $z_i$, given the input sequence $x$ and all preceding tokens $z_{<i}$.
\begin{figure*}[t]
  \centering
  \includegraphics[width=\linewidth]{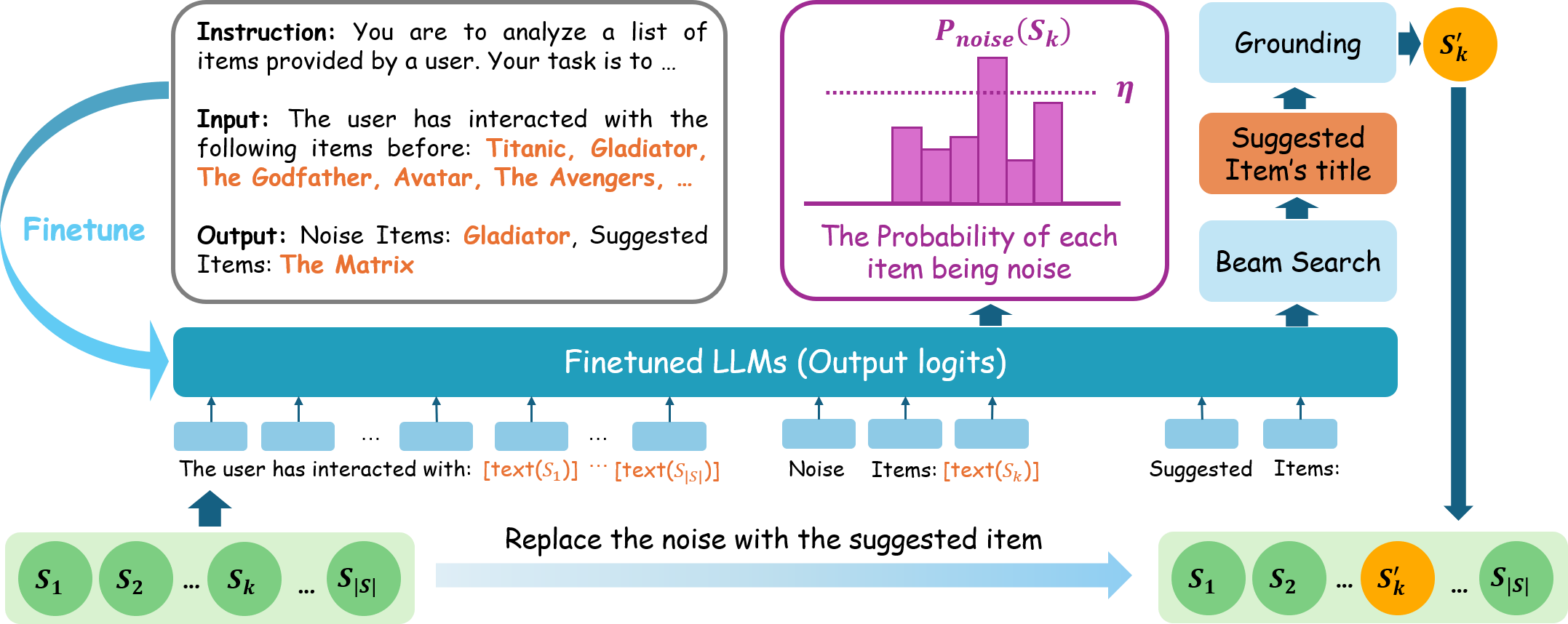}
  \caption{Schematic diagram of the LLM4DSR.}
  \Description{}
  \label{method}
\end{figure*}

\subsection{Uncertainty Estimation Module}
\label{uncertainty_estimation}
While the aforementioned instruction tuning has activated the denoising potential of LLMs, they still face two challenges during applications: 1) Given the notorious hallucination phenomenon inherent in LLMs and the complexity of the denoising task, directly relying on the model's response to identify noise items might not be reliable or precise. The model may generate fake items that are not actually present in the sequence, or it may lack high confidence in the given item, thereby compromising the accuracy of the denoising results. 2) The response is limited to addressing a single noisy item per sequence.  This is because the training data used during fine-tuning contains only one noisy item in the output, leading the model to consistently output a single item during inference \footnote{We also have attempted to construct instruction data with multiple noisy items and guide the LLM to identify several noise items. However, this approach resulted in very low accuracy. We hypothesize that this is due to the increased difficulty of directly identifying a varying number of noisy items.}. Consequently, this strategy is inflexible in handling cases where multiple noisy items are present.

To tackle these issues, we introduce an uncertainty estimation strategy. Rather than directly utilizing the response from LLMs, which could be imprecise and inflexible, we opt to scrutinize the probability of each item  $s_k$ in the sequence $S$ being regarded as a noise item by LLMs:
\begin{equation}
P_{\text{noise}}(s_k) = P(z_{\text{noise}}= T_k | x) = \prod \limits_{j=1}^{|T_k|} P(T_{k,j} | x, T_{k,<j})
\end{equation}
where we examine the generative probability of the description of the item $s_k$. Here, $z_{\text{noise}}$ represents the token sequence of the noise item. $T_k$ denotes the token sequence of $text(s_k)$, $|T_k|$ denotes the total length of \( T_k\) and $T_{k,j}$ represents the $j$-th token of \( T_k \). 

Considering the calculation of $P_{noise}(s_k)$ can be time-consuming due to the need to iterate over each token, in practice, we may simplify the process by calculating the probability of only the first token. This is because the first token of the items in a sequence is typically unique, suggesting that the generated item description can be determined primarily by the first token. Our empirical analysis also demonstrates that the generative probability for the remaining tokens is approximately 1.

\begin{figure}[t]
  \centering
  \includegraphics[width=0.8\linewidth]{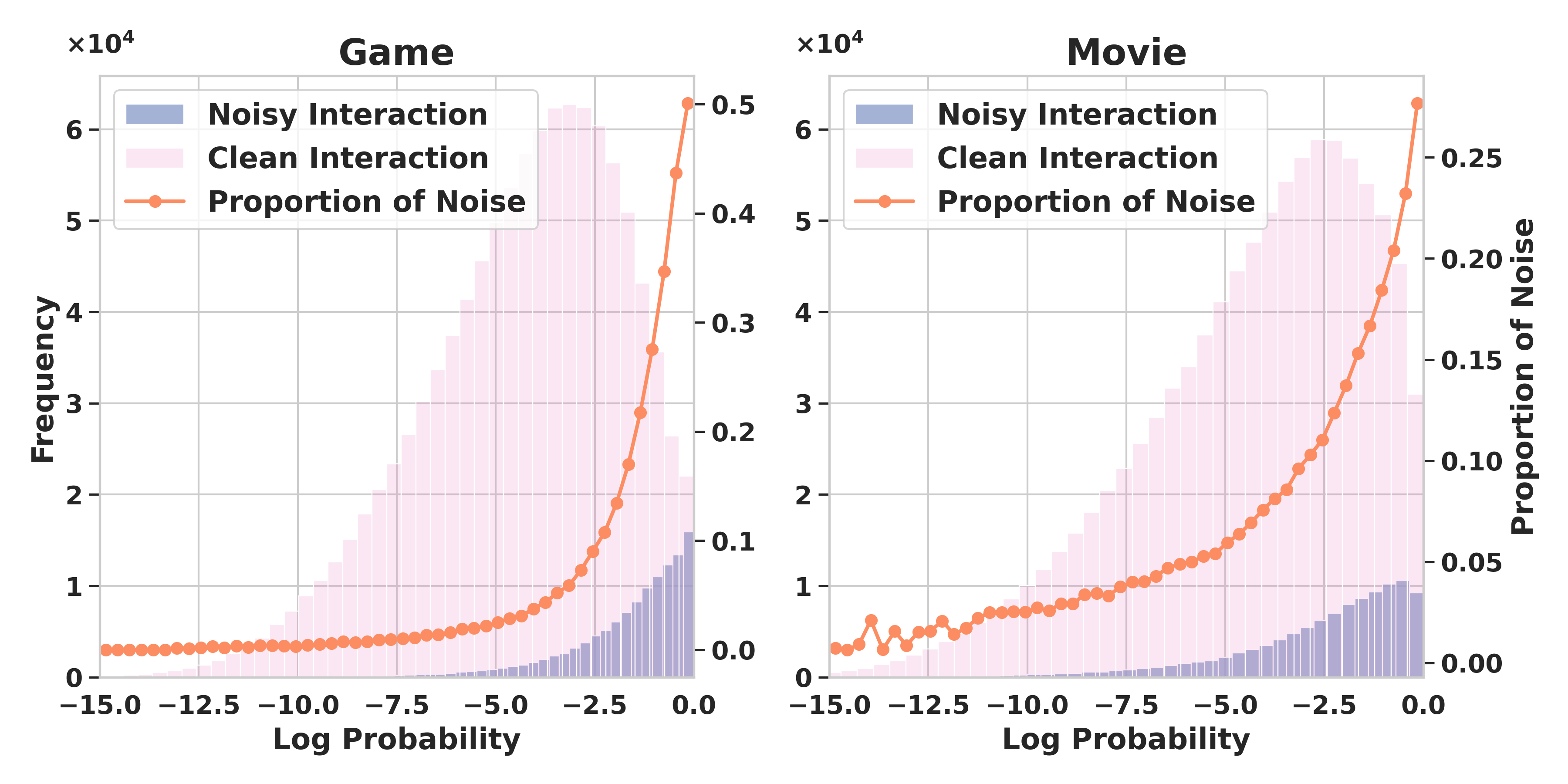}
  \caption{The distribution of the logarithmic probability of the model-generated samples being noise.}
  \Description{}
  \label{noise_prob_distribution}
\end{figure}

The probability $P_{noise}(s_k)$ reflects the confidence of the LLM's judgment --- the higher the probability, the more likely the item is to be a noisy item. To empirically validate its effectiveness, we conducted experiments on a dataset with an artificial noise ratio of 10\% \footnote{We use a dataset with artificially injected noise because only in this setting the ground-truth labels for noisy items are available.}. The details of the dataset are provided in Section \ref{exp_dataset}. As shown in Figure \ref{noise_prob_distribution}, as the probability $P_{noise}(s_k)$ increases, both the number and proportion of noise samples increase, while the number of clean samples decreases.

Thus, in our approach, $P_{noise}(s_k)$ is used to identify noisy items within the sequence. Formally, we apply a threshold $\eta$ for classification:
\begin{equation}
y_k = \begin{cases}
1, &  P_{noise}(s_k) > \eta \\
0, &  P_{noise}(s_k) \leq  \eta 
\end{cases}
\end{equation}

This uncertainty-based strategy mitigates the hallucination issue inherent in LLMs by avoiding reliance on potentially erroneous generative outputs. Besides, this strategy provides flexibility in handling multiple noisy items, as the model can assign a high $P_{noise}(s_k)$ to multiple items in the sequence.


\subsection{Replace Noise with Suggested Items}
\label{replace}
Beyond removing noise items, we propose leveraging the generative capabilities of LLMs to suggest replacement items. The replacement items could bring additional knowledge of user preferences, potentially leading to improved recommendation performance. Specifically, for each identified noisy item $s_k$, we use the prompt \textit{\textless User Interactions, Noise Items: $text(s_k)$, Suggested Items: \textgreater} to guide the LLM in generating a description of the suggested replacement item $text_{suggest}$. 
Given the hallucination issue of LLMs, $text_{suggest}$ could be a fake item that does not necessarily exist in the item set. Grounding techniques \cite{bao2023bi} can be applied to find the most close existing item:
\begin{equation}
s_k' = \argmin_{v_i \in \mathcal{V}} f(text_{suggest}, text(v_i))
\end{equation}
where $f(.,.)$ represents the similarity between two textual descriptions, which can be implemented via the embedding similarity of texts. The item $s_k'$ is then used to replace $s_k$, generating a corrected sequence that can be leveraged to enhance the performance of various recommendation models.

\section{Experiments}
\label{Experiments}
We aim to answer the following research questions:
\begin{itemize}[leftmargin=*]
\item \textbf{RQ1:} How does LLM4DSR perforem compare to existing state-of-the-art sequential recommendation denoising methods?
\item \textbf{RQ2:} What are the impacts of the components (e.g., uncertainty estimation module, replace with suggested items) on LLM4DSR?
\item \textbf{RQ3:} How does hyperparameter $\eta$ affect model performance?
\item \textbf{RQ4:} How does LLM4DSR identify noise and complete the sequence?
\end{itemize}

\subsection{Experimental Settings}
\subsubsection{Datasets.}
\label{exp_dataset}
Three conventional real-world datasets: \textit{Amazon Video Games}, \textit{Amazon Toy and Games}, and \textit{Amazon Movies} \footnote{\url{https://jmcauley.ucsd.edu/data/amazon/index_2014.html}} were utilized in our experiments. These datasets are commonly used for the studies of LLM-based sequential recommendation \cite{cui2024distillation, bao2023bi, cao2024aligning, lee2024star}. To ensure a fair comparison, we adopted the same data preprocessing used in recent studies \cite{bao2023bi}. That is, for each user sequence longer than 11 interactions, a sliding window of length 11 is used to segment the sequences. The sequences were then organized in ascending order of timestamps to partition each dataset into training, validation, and testing sets with ratios of 8:1:1. We randomly retained 20,000 items for \textit{Amazon Movies} due to its large size. The dataset statistics are presented in Table \ref{tab:statistics}.

We employ two noise settings in our experiments: (1) the natural noise setting, which directly uses the original real-world datasets, as these datasets inherently contain noise introduced during practical data collection; and (2) the artificial noise setting, where a certain ratio of noise, denoted by $\alpha$, is explicitly injected into the dataset. Specifically, in the artificial noise setting, each item in a sequence has a probability of $\alpha$ of being replaced by a randomly selected noise item. This setting allows us to evaluate the effectiveness of the denoising strategy under varying levels of noise. Moreover, the artificial noise setting provides ground-truth labels for the noise items, which can be used to assess the accuracy of different methods in identifying noise items. Note that the artificial noise is introduced before segmenting the sequences using the sliding window, thus preventing any information leakage between sequences.


\begin{table}[t]
\centering
\caption{Statistics of the datasets.}
\label{tab:statistics}
\resizebox{0.7\columnwidth}{!}{%
\begin{tabular}{@{}lccccc@{}}
\toprule
\textbf{Datasets} & \textbf{\#Users} & \textbf{\#Items} & \textbf{\#Interactions} & \textbf{\#Sequences} & \textbf{\#Density} \\ \midrule
Games              & 54955            & 17256            & 494934                  & 149393               & 0.0522\%           \\
Toy               & 19124            & 11758            & 165247                  & 47931                & 0.0735\%           \\
Movie             & 9329             & 5101             & 214980                  & 121690               & 0.4518\%           \\ \bottomrule
\end{tabular}%
}
\end{table}

\subsubsection{Baselines.}
We conducted experiments on three backbones, including the traditional sequential recommendation models SASRec \cite{kang2018self} and BERT4Rec \cite{sun2019bert4rec}, as well as a LLM-based recommendation model LLaRA \cite{liao2024llara}. We compare LLM4DSR with the following denoising baselines \footnote{We acknowledge the existence of other explicit denoising methods for sequential recommendation, such as SSDRec \cite{zhang2024ssdrec}, END4Rec \cite{han2024end4rec}. We excluded them from our study because these methods are not compatible with the setting of this paper.}:
\begin{itemize}
    \item \textbf{HSD} \cite{zhang2022hierarchical} is an explicit denoising method. It posits that noisy interactions within a sequence do not align with the user's long-term and short-term interests and exhibit noticeable discontinuities with the sequence context. It jointly identifies noise items in the sequence from both user-level and sequence-level signals.
    \item \textbf{STEAM} \cite{lin2023self} is an explicit denoising method. It constructs a self-supervised task by randomly replacing items in the original sequence, training a discriminator and a generator to determine whether an item in the sequence is noise and to complete the sequence, respectively.
    \item \textbf{FMLP} \cite{zhou2022filter} is an implicit denoising method. It uses Fast Fourier Transform and a learnable filter to eliminate noise in the representation.
    \item \textbf{CL4SRec} \cite{xie2022contrastive} is an implicit denoising method. It constructs positive samples of the original sequence through transformations such as item cropping, item masking, and item reordering and then uses contrastive learning to enhance the model's robustness to noise in the sequence.
\end{itemize}
We tested implicit denoising methods only on traditional sequential recommendation backbones. This is due to the coupling of these methods with specific models, which prevents their application to LLM-based recommendation systems.


\subsubsection{Evaluation Metrics.}
We utilize the following metrics to evaluate our method. The evaluation metrics employed were NDCG@$K$ and Hit Rate@$K$, with $K$ set to 20 and 50.:
\begin{itemize}
    \item \textbf{Hit Ratio@K} measures the proportion of positive items that appear in the top K positions of the recommendation list.
    \begin{equation}
    \begin{aligned}
    & \mathrm{HR}_u @ K=\frac{1}{|\mathcal{P}(u)|} \sum_{i \in \mathcal{P}(u)} \mathrm{I}\left[q_{u i} \leq K\right] \\
    & \mathrm{HR} @ K=\frac{1}{|\mathcal{U}|} \sum_{u \in \mathcal{U}} \mathrm{HR}_u @ K
    \end{aligned}
    \end{equation}
    where $\mathrm{I}[]$ denotes an indicator function; $\mathcal P(u)$ denotes the positive item set in the test data for a user $u$; $q_{ui}$ denotes the ranking position of the item $i$ for the user $u$.
    \item \textbf{NDCG@K} measures the ranking quality of recommendation through discounted importance based on the position:
    \begin{equation}
    \begin{aligned}
    &DCG_u@K=\sum_{i \in \mathcal P(u)} \frac{\mathrm{I}[q_{ui} \leq K]}{\log \left(q_{ui}+1\right)} \\
    & ND C G @ K=\frac{1}{|\mathcal{U}|} \sum_{u \in \mathcal{U}} \frac{D C G_u @ K}{I D C G_u @ K}
    \end{aligned}
    \end{equation}
     $IDCG_u@K$ is the $DCG_u@K$ value of the ideal ranking with the optimal ranking for the user $u$.
\end{itemize}

\subsubsection{Implementation Details.}
For training LLM4DSR, we select Llama3-8B \cite{dubey2024llama} as the backbone and randomly sample 5000 prompts to fine-tune the model over 50 epochs using LoRA \cite{hu2021lora}. For the threshold hyperparameter \(\eta\), we conduct a search within the [0,1] quantiles of the noise probabilities across all items, with a search interval of 0.05. In the experiment, we found that \(\eta\) demonstrated strong robustness. Searching above the 0.7 quantile yielded satisfying results. A sensitivity analysis of this parameter is provided in Section \ref{sec:train_noise_threshold}. Additionally, to prevent excessive modification of any sequence, we limit each sequence to be modified by at most $b$ samples where the parameter $b$ is chosen from $\{1,2,3\}$.

For SASRec and BERT4Rec, we fix the learning rate at 0.001 and the embedding size at 64. For the LLM-based recommendation model, LLaRA, the original configuration involves selecting the Top 1 item from the candidate items as the recommendation result, without item ranking process, which prevents the calculation of previously introduced evaluation metrics. To align with standard experimental settings, we follow \cite{bao2023bi} and implement a grounding operation to enable LLaRA rank the entire item set.

We use the source code provided in the original papers and search for optimal hyperparameters for all comparison methods according to the instructions in the original papers. 
Besides, for the two explicit denoising methods, HSD and STEAM, we performe end-to-end training on the backbone used in the original paper (i.e., BERT4Rec) and generate the denoised dataset. To facilitate comparison with other backbones (i.e., SASRec and LLaRA), we transferred the denoised dataset to these backbones and trained them from scratch.

\begin{table}[]
\centering
\caption{The NDCG performance comparison on the natural noise setting using different backbone. The best result is bolded and the runner-up is underlined.}
\label{tab:cp_real_world_ndcg}
\resizebox{\textwidth}{!}{%
\begin{tabular}{@{}l|l|cc|cc|cc@{}}
\toprule
\multicolumn{1}{c|}{\multirow{2}{*}{Backbone}} & \multicolumn{1}{c|}{\multirow{2}{*}{Method}} & \multicolumn{2}{c|}{Games}        & \multicolumn{2}{c|}{Toy}          & \multicolumn{2}{c}{Movie}         \\ \cmidrule(l){3-8} 
\multicolumn{1}{c|}{}                          & \multicolumn{1}{c|}{}                        & NDCG@20         & NDCG@50         & NDCG@20         & NDCG@50         & NDCG@20         & NDCG@50         \\ \midrule
\multirow{6}{*}{SASRec}                        & None                                         & 0.0299          & 0.0377          & 0.0190          & 0.0239          & 0.0468          & 0.0548          \\ \cmidrule(lr){2-2}
                                               & FMLP                                         & 0.0297          & 0.0377          & 0.0200          & 0.0245          & 0.0488          & 0.0574          \\
                                               & CL4SRec                                      & \textbf{0.0331} & \textbf{0.0405} & {\ul 0.0201}    & {\ul 0.0247}    & {\ul 0.0492}    & {\ul 0.0579}    \\ \cmidrule(lr){2-2}
                                               & HSD                                          & 0.0246          & 0.0324          & 0.0121          & 0.0172          & 0.0369          & 0.0451          \\
                                               & STEAM                                        & 0.0233          & 0.0325          & 0.0187          & 0.0242          & 0.0475          & 0.0569          \\ \cmidrule(lr){2-2}
                                               & LLM4SRD                                      & {\ul 0.0315}    & {\ul 0.0404}    & \textbf{0.0208} & \textbf{0.0262} & \textbf{0.0505} & \textbf{0.0597} \\ \midrule
\multirow{6}{*}{BERT4Rec}                      & None                                         & 0.0351          & 0.0436          & {\ul 0.0205}    & {\ul 0.0256}    & 0.0584          & {\ul 0.0684}    \\ \cmidrule(lr){2-2}
                                               & FMLP                                         & 0.0369          & 0.0449          & 0.0195          & 0.0242          & 0.0554          & 0.0641          \\
                                               & CL4SRec                                      & 0.0359          & 0.0438          & 0.0192          & 0.0246          & {\ul 0.0587}    & 0.0676          \\ \cmidrule(lr){2-2}
                                               & HSD                                          & 0.0301          & 0.0387          & 0.0187          & 0.0230          & 0.0467          & 0.0565          \\
                                               & STEAM                                        & {\ul 0.0373}    & {\ul 0.0453}    & 0.0195          & 0.0250          & 0.0548          & 0.0643          \\ \cmidrule(lr){2-2}
                                               & LLM4SRD                                      & \textbf{0.0375} & \textbf{0.0464} & \textbf{0.0222} & \textbf{0.0280} & \textbf{0.0604} & \textbf{0.0697} \\ \midrule
\multirow{4}{*}{LLaRA}                         & None                                         & {\ul 0.0261}    & {\ul 0.0341}    & {\ul 0.0224}    & {\ul 0.0275}    & 0.0388          & 0.0472          \\ \cmidrule(lr){2-2}
                                               & HSD                                          & 0.0231          & 0.0305          & 0.0136          & 0.0188          & 0.0271          & 0.0347          \\
                                               & STEAM                                        & 0.0200          & 0.0284          & 0.0182          & 0.0244          & {\ul 0.0396}    & {\ul 0.0488}    \\ \cmidrule(lr){2-2}
                                               & LLM4SRD                                      & \textbf{0.0290} & \textbf{0.0375} & \textbf{0.0241} & \textbf{0.0306} & \textbf{0.0409} & \textbf{0.0504} \\ \bottomrule
\end{tabular}%
}
\end{table}

\begin{table}[]
\centering
\caption{The Hit ratio performance comparison on the natural noise setting using different backbone. The best result is bolded and the runner-up is underlined.}
\label{tab:cp_real_world_hit}
\resizebox{0.8\textwidth}{!}{%
\begin{tabular}{@{}l|l|cc|cc|cc@{}}
\toprule
\multicolumn{1}{c|}{\multirow{2}{*}{Backbone}} & \multicolumn{1}{c|}{\multirow{2}{*}{Method}} & \multicolumn{2}{c|}{Games}        & \multicolumn{2}{c|}{Toy}          & \multicolumn{2}{c}{Movie}         \\ \cmidrule(l){3-8} 
\multicolumn{1}{c|}{}                          & \multicolumn{1}{c|}{}                        & HR@20           & HR@50           & HR@20           & HR@50           & HR@20           & HR@50           \\ \midrule
\multirow{6}{*}{SASRec}                        & None                                         & 0.0646          & 0.1040          & 0.0501          & 0.0749          & 0.0900          & 0.1308          \\ \cmidrule(lr){2-2}
                                               & FMLP                                         & 0.0600          & 0.1004          & {\ul 0.0532}    & 0.0761          & 0.0924          & 0.1360          \\
                                               & CL4SRec                                      & {\ul 0.0696}    & {\ul 0.1072}    & 0.0515          & 0.0751          & {\ul 0.0958}    & {\ul 0.1400}    \\ \cmidrule(lr){2-2}
                                               & HSD                                          & 0.0514          & 0.0908          & 0.0319          & 0.0582          & 0.0780          & 0.1196          \\
                                               & STEAM                                        & 0.0578          & 0.1052          & 0.0501          & {\ul 0.0778}    & 0.0906          & 0.1386          \\ \cmidrule(lr){2-2}
                                               & LLM4SRD                                      & \textbf{0.0696} & \textbf{0.1148} & \textbf{0.0557} & \textbf{0.0830} & \textbf{0.0976} & \textbf{0.1440} \\ \midrule
\multirow{6}{*}{BERT4Rec}                      & None                                         & 0.0688          & 0.1120          & 0.0461          & 0.0718          & 0.0982          & {\ul 0.1484}    \\ \cmidrule(lr){2-2}
                                               & FMLP                                         & 0.0710          & 0.1116          & 0.0426          & 0.0699          & 0.0952          & 0.1394          \\
                                               & CL4SRec                                      & 0.0692          & 0.1088          & {\ul 0.0469}    & {\ul 0.0738}    & {\ul 0.1022}    & 0.1476          \\ \cmidrule(lr){2-2}
                                               & HSD                                          & 0.0708          & {\ul 0.1144}    & 0.0430          & 0.0651          & 0.0802          & 0.1298          \\
                                               & STEAM                                        & {\ul 0.0714}    & 0.1122          & 0.0446          & 0.0724          & 0.0932          & 0.1414          \\ \cmidrule(lr){2-2}
                                               & LLM4SRD                                      & \textbf{0.0748} & \textbf{0.1202} & \textbf{0.0505} & \textbf{0.0799} & \textbf{0.1046} & \textbf{0.1516} \\ \midrule
\multirow{4}{*}{LLaRA}                         & None                                         & 0.0512          & 0.0918          & {\ul 0.0538}    & {\ul 0.0797}    & 0.0786          & 0.1208          \\ \cmidrule(lr){2-2}
                                               & HSD                                          & 0.0490          & 0.0866          & 0.0357          & 0.0617          & 0.0688          & 0.1072          \\
                                               & STEAM                                        & {\ul 0.0554}    & {\ul 0.0990}    & 0.0474          & 0.0786          & {\ul 0.0820}    & {\ul 0.1288}    \\ \cmidrule(lr){2-2}
                                               & LLM4SRD                                      & \textbf{0.0602} & \textbf{0.1026} & \textbf{0.0576} & \textbf{0.0905} & \textbf{0.0870} & \textbf{0.1350} \\ \bottomrule
\end{tabular}%
}
\end{table}

\begin{table}[]
\centering
\caption{The NDCG performance comparison on the artificial noise setting using different backbone. The noise ratio is 10\%. The best result is bolded and the runner-up is underlined.}
\label{tab:cp_artificial_ndcg}
\resizebox{\textwidth}{!}{%
\begin{tabular}{@{}l|l|cc|cc|cc@{}}
\toprule
\multicolumn{1}{c|}{\multirow{2}{*}{Backbone}} & \multicolumn{1}{c|}{\multirow{2}{*}{Method}} & \multicolumn{2}{c|}{Games}        & \multicolumn{2}{c|}{Toy}          & \multicolumn{2}{c}{Movie}         \\ \cmidrule(l){3-8} 
\multicolumn{1}{c|}{}                          & \multicolumn{1}{c|}{}                        & NDCG@20         & NDCG@50         & NDCG@20         & NDCG@50         & NDCG@20         & NDCG@50         \\ \midrule
\multirow{6}{*}{SASRec}                        & None                                         & 0.0252          & 0.0314          & 0.0139          & 0.0178          & 0.0390          & 0.0455          \\ \cmidrule(lr){2-2}
                                               & FMLP                                         & 0.0253          & 0.0315          & 0.0138          & 0.0173          & {\ul 0.0415}    & {\ul 0.0484}    \\
                                               & CL4SRec                                      & {\ul 0.0278}    & {\ul 0.0336}    & {\ul 0.0151}    & {\ul 0.0183}    & 0.0405          & 0.0464          \\ \cmidrule(lr){2-2}
                                               & HSD                                          & 0.0198          & 0.0250          & 0.0090          & 0.0127          & 0.0340          & 0.0407          \\
                                               & STEAM                                        & 0.0261          & 0.0333          & 0.0123          & 0.0161          & 0.0401          & 0.0476          \\ \cmidrule(lr){2-2}
                                               & LLM4SRD                                      & \textbf{0.0292} & \textbf{0.0357} & \textbf{0.0152} & \textbf{0.0200} & \textbf{0.0431} & \textbf{0.0517} \\ \midrule
\multirow{6}{*}{BERT4Rec}                      & None                                         & 0.0266          & 0.0335          & 0.0150          & 0.0195          & 0.0485          & 0.0569          \\ \cmidrule(lr){2-2}
                                               & FMLP                                         & 0.0281          & 0.0353          & 0.0147          & 0.0188          & 0.0478          & 0.0561          \\
                                               & CL4SRec                                      & 0.0268          & 0.0341          & 0.0153          & 0.0196          & {\ul 0.0498}    & {\ul 0.0579}    \\ \cmidrule(lr){2-2}
                                               & HSD                                          & 0.0288          & 0.0361          & 0.0152          & 0.0189          & 0.0390          & 0.0459          \\
                                               & STEAM                                        & {\ul 0.0294}    & {\ul 0.0367}    & {\ul 0.0182}    & {\ul 0.0218}    & 0.0472          & 0.0544          \\ \cmidrule(lr){2-2}
                                               & LLM4SRD                                      & \textbf{0.0316} & \textbf{0.0397} & \textbf{0.0183} & \textbf{0.0231} & \textbf{0.0509} & \textbf{0.0591} \\ \midrule
\multirow{4}{*}{LLaRA}                         & None                                         & 0.0202          & 0.0264          & {\ul 0.0152}    & {\ul 0.0195}    & 0.0323          & 0.0387          \\ \cmidrule(lr){2-2}
                                               & HSD                                          & 0.0146          & 0.0198          & 0.0116          & 0.0153          & 0.0197          & 0.0267          \\
                                               & STEAM                                        & {\ul 0.0207}    & {\ul 0.0275}    & 0.0144          & 0.0191          & {\ul 0.0325}    & {\ul 0.0395}    \\ \cmidrule(lr){2-2}
                                               & LLM4SRD                                      & \textbf{0.0252} & \textbf{0.0321} & \textbf{0.0150} & \textbf{0.0204} & \textbf{0.0358} & \textbf{0.0445} \\ \bottomrule
\end{tabular}%
}
\end{table}

\begin{table}[]
\centering
\caption{The Hit ratio performance comparison on the artificial noise setting using different backbone. The noise ratio is 10\%. The best result is bolded and the runner-up is underlined.}
\label{tab:cp_artificial_hit}
\resizebox{0.8\textwidth}{!}{%
\begin{tabular}{@{}l|l|cc|cc|cc@{}}
\toprule
\multicolumn{1}{c|}{\multirow{2}{*}{Backbone}} & \multicolumn{1}{c|}{\multirow{2}{*}{Method}} & \multicolumn{2}{c|}{Games}        & \multicolumn{2}{c|}{Toy}          & \multicolumn{2}{c}{Movie}         \\ \cmidrule(l){3-8} 
\multicolumn{1}{c|}{}                          & \multicolumn{1}{c|}{}                        & HR@20           & HR@50           & HR@20           & HR@50           & HR@20           & HR@50           \\ \midrule
\multirow{6}{*}{SASRec}                        & None                                         & 0.0526          & 0.0838          & 0.0348          & 0.0542          & 0.0750          & 0.1084          \\ \cmidrule(lr){2-2}
                                               & FMLP                                         & 0.0508          & 0.0824          & 0.0365          & 0.0547          & {\ul 0.0772}    & {\ul 0.1120}    \\
                                               & CL4SRec                                      & {\ul 0.0556}    & 0.0864          & {\ul 0.0394}    & {\ul 0.0557}    & 0.0770          & 0.1070          \\ \cmidrule(lr){2-2}
                                               & HSD                                          & 0.0412          & 0.0678          & 0.0254          & 0.0444          & 0.0700          & 0.1042          \\
                                               & STEAM                                        & 0.0532          & {\ul 0.0900}    & 0.0321          & 0.0513          & 0.0740          & 0.1118          \\ \cmidrule(lr){2-2}
                                               & LLM4SRD                                      & \textbf{0.0624} & \textbf{0.0956} & \textbf{0.0401} & \textbf{0.0642} & \textbf{0.0876} & \textbf{0.1314} \\ \midrule
\multirow{6}{*}{BERT4Rec}                      & None                                         & 0.0534          & 0.0880          & 0.0346          & 0.0574          & 0.0826          & 0.1258          \\ \cmidrule(lr){2-2}
                                               & FMLP                                         & 0.0598          & 0.0962          & 0.0330          & 0.0536          & 0.0818          & 0.1240          \\
                                               & CL4SRec                                      & 0.0538          & 0.0906          & 0.0369          & {\ul 0.0586}    & {\ul 0.0864}    & {\ul 0.1276}    \\ \cmidrule(lr){2-2}
                                               & HSD                                          & {\ul 0.0600}    & {\ul 0.0970}    & 0.0340          & 0.0530          & 0.0682          & 0.1032          \\
                                               & STEAM                                        & 0.0562          & 0.0936          & {\ul 0.0388}    & 0.0572          & 0.0836          & 0.1202          \\ \cmidrule(lr){2-2}
                                               & LLM4SRD                                      & \textbf{0.0644} & \textbf{0.1056} & \textbf{0.0401} & \textbf{0.0645} & \textbf{0.0890} & \textbf{0.1306} \\ \midrule
\multirow{4}{*}{LLaRA}                         & None                                         & 0.0466          & 0.0784          & {\ul 0.0363}    & 0.0582          & {\ul 0.0668}    & 0.0992          \\ \cmidrule(lr){2-2}
                                               & HSD                                          & 0.0332          & 0.0598          & 0.0273          & 0.0459          & 0.0482          & 0.0842          \\
                                               & STEAM                                        & {\ul 0.0476}    & {\ul 0.0822}    & 0.0361          & {\ul 0.0599}    & 0.0664          & {\ul 0.1018}    \\ \cmidrule(lr){2-2}
                                               & LLM4SRD                                      & \textbf{0.0514} & \textbf{0.0866} & \textbf{0.0380} & \textbf{0.0655} & \textbf{0.0770} & \textbf{0.1216} \\ \bottomrule
\end{tabular}%
}
\end{table}

\subsection{Performance Comparison (RQ1)}
In this section, we analyze the superior of LLM4DSR under two different noise settings as compared with other baselines.
\subsubsection{Evaluations on the nature noise setting.}
In Table \ref{tab:cp_real_world_ndcg} and Table \ref{tab:cp_real_world_hit}, we present the test results of various methods on real-world datasets. Notably, LLM4DSR consistently achieves the best performance across almost all datasets. 

\textbf{For explicit denoising methods (\ie STEAM and HSD)}, we observed that while these methods perform well on BERT4Rec, their effectiveness diminishes when retraining on SASRec using the denoised dataset they generated. For instance, on Amazon Games, STEAM improves NDCG@20 from 0.0351 to 0.0373 on BERT4Rec. However, when the same denoised dataset is evaluated on SASRec, NDCG@20 drops from 0.0299 to 0.0233. This suggests that the denoised dataset generated by STEAM lacks good transferability. Additionally, the performance of HSD significantly declined. We argue that this issue may arise from HSD's inability to effectively capture sequence information in short sequence scenarios. This limitation can result in the loss of critical information when incorrect items are deleted. In contrast, LLM4DSR demonstrates consistently superior performance across all datasets and backbones, proving its ability to effectively identify and replace noisy samples based on the internal knowledge of LLMs.

\textbf{For implicit denoising methods (\ie FMLP and CL4SRec)}, we found that their performance is unstable and can even decline on certain datasets. For example, on the Toy dataset, both FMLP and CL4SRec experienced performance drops, with NDCG@20 decreasing from 0.256 to 0.242 and 0.246, respectively. We hypothesize that for CL4SRec, the substantial difference from the original sequence during the sequence enhancement phase may degrade the quality of positive samples, and using contrastive loss for alignment might introduce additional noise. FMLP, which performs denoising at the representation level through filtering, appears unable to effectively mitigate the impact of noise samples within the sequence.

\subsubsection{Evaluations on the artifically noise setting.}
\begin{figure}[t]
  \centering
  \includegraphics[width=\linewidth]{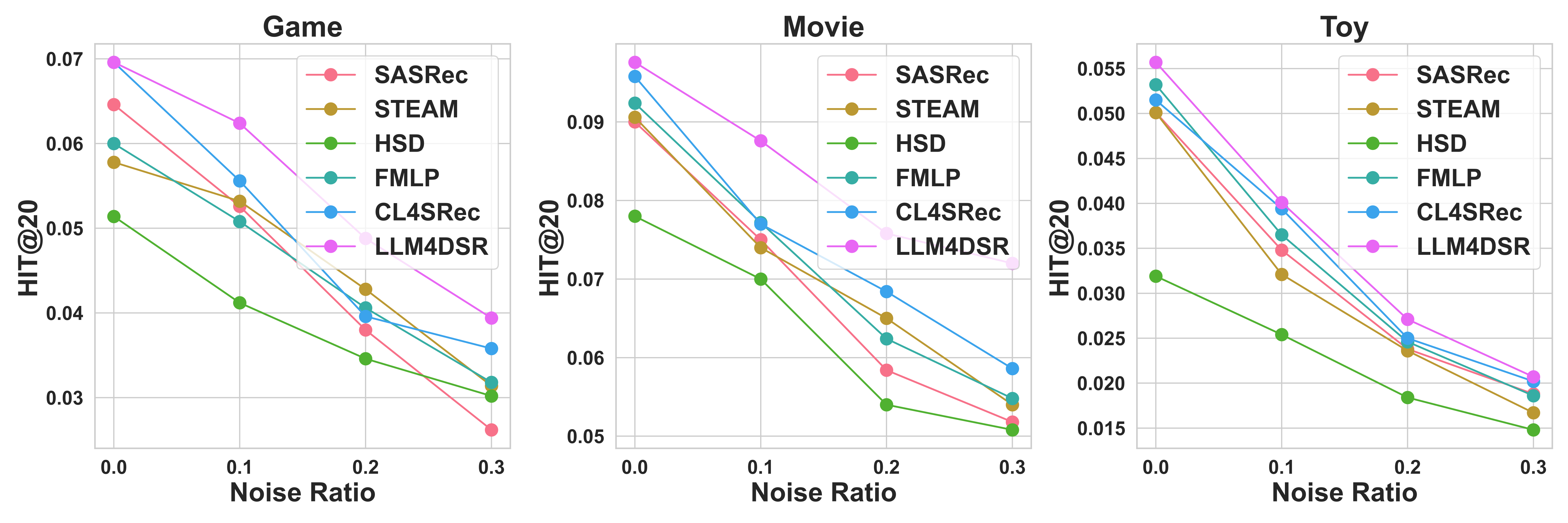}
  \caption{The performance of different methods on the artifically noise setting with varying noise ratios from 0\% to 30\%. SASRec serves as the backbone model. A noise ratio of 0.0 represents a real-world dataset containing natural noise.}
  \Description{}
  \label{cp_noise_ratio}
\end{figure}

The test results for various methods on the artifically noise setting, with a noise ratio of 10\%, are presented in Table \ref{tab:cp_artificial_ndcg} and Table \ref{tab:cp_artificial_hit}. The results indicate that LLM4DSR consistently outperforms all comparative methods. Artificial noise, being more distinguishable than natural noise in conventional real-world datasets, makes the improvement of LLM4DSR more pronounced on such datasets. For example, on Amazon Games, LLM4DSR combined with SASRec achieves a 5.3\% improvement in NDCG@20, whereas on the artificially noise-added dataset, the improvement reaches 15.8\%. Furthermore, we evaluated the performance of all comparative methods under different noise ratios, as illustrated in Figure \ref{cp_noise_ratio}. LLM4DSR significantly surpasses other comparative methods demonstrate its robust performance across varying noise conditions.

\subsubsection{Evaluation of the accuracy of explicit denoising methods in identifying artificial noise.}
\begin{figure}[t]
  \centering
  \includegraphics[width=0.8\linewidth]{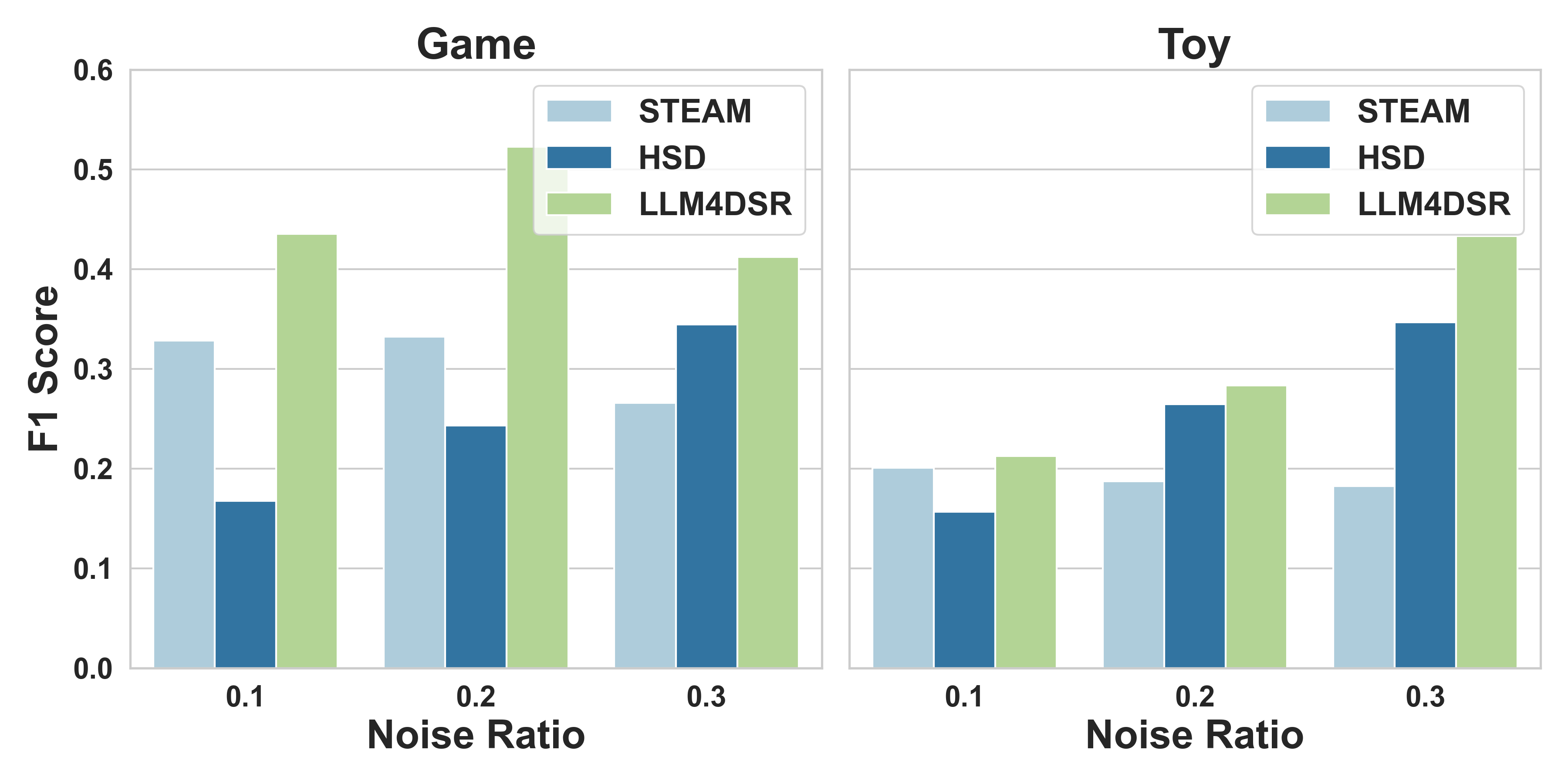}
  \caption{F1 score of explicit denoising method for noise recognition.}
  \Description{}
  \label{F1_score}
\end{figure}

The explicit denoising method achieves denoising by identifying and removing noise, with the accuracy of noise identification being closely related to its overall effectiveness. To better understand the performance of explicit denoising methods, we evaluated the F1 score of various explicit denoising methods on datasets with artificially introduced noise, where noise labels are available. As illustrated in Figure \ref{F1_score}, we compare LLM4DSR with two other state-of-the-art explicit denoising methods on the Game and Toy datasets with different noise ratios of 10\%, 20\%, and 30\%. Among these methods, LLM4DSR consistently achieved the highest performance, suggesting that leveraging the open knowledge of LLM can significantly enhance noise identification.

\subsection{Ablation Study (RQ2)}
\subsubsection{The impact of different components in LLM4DSR}
To verify the contributions of various components of LLM4DSR, we conducted an ablation study on different variants under SASRec, using both real and artificially noisy datasets from the Game and Toy datasets. The variants tested include: (1) "w/o uncertainty," which omits uncertainty estimation and directly uses the LLM output for denoising; (2) "w/o completion," which excludes the completion capability of LLM4DSR and performs only deletion operations on sequences; and (3) "w/ popularity completion," which replaces noise with items randomly sampled from the item set based on popularity weights. As presented in Table \ref{tab:ablation_study2}, the consistently superior performance of LLM4DSR across all datasets highlights the effectiveness of each component. In particular, the performance decline observed in the "w/o completion" and "w/ popularity completion" highlights the critical role of the replacement operation. This demonstrates that LLM4DSR effectively extracts user interests and suggests items that better align with the sequence, while also mitigating the negative impact of erroneous deletions.

\begin{figure}[t]
  \centering
  \abovecaptionskip=-0pt
  \belowcaptionskip=-10pt
  \includegraphics[width=0.8\linewidth]{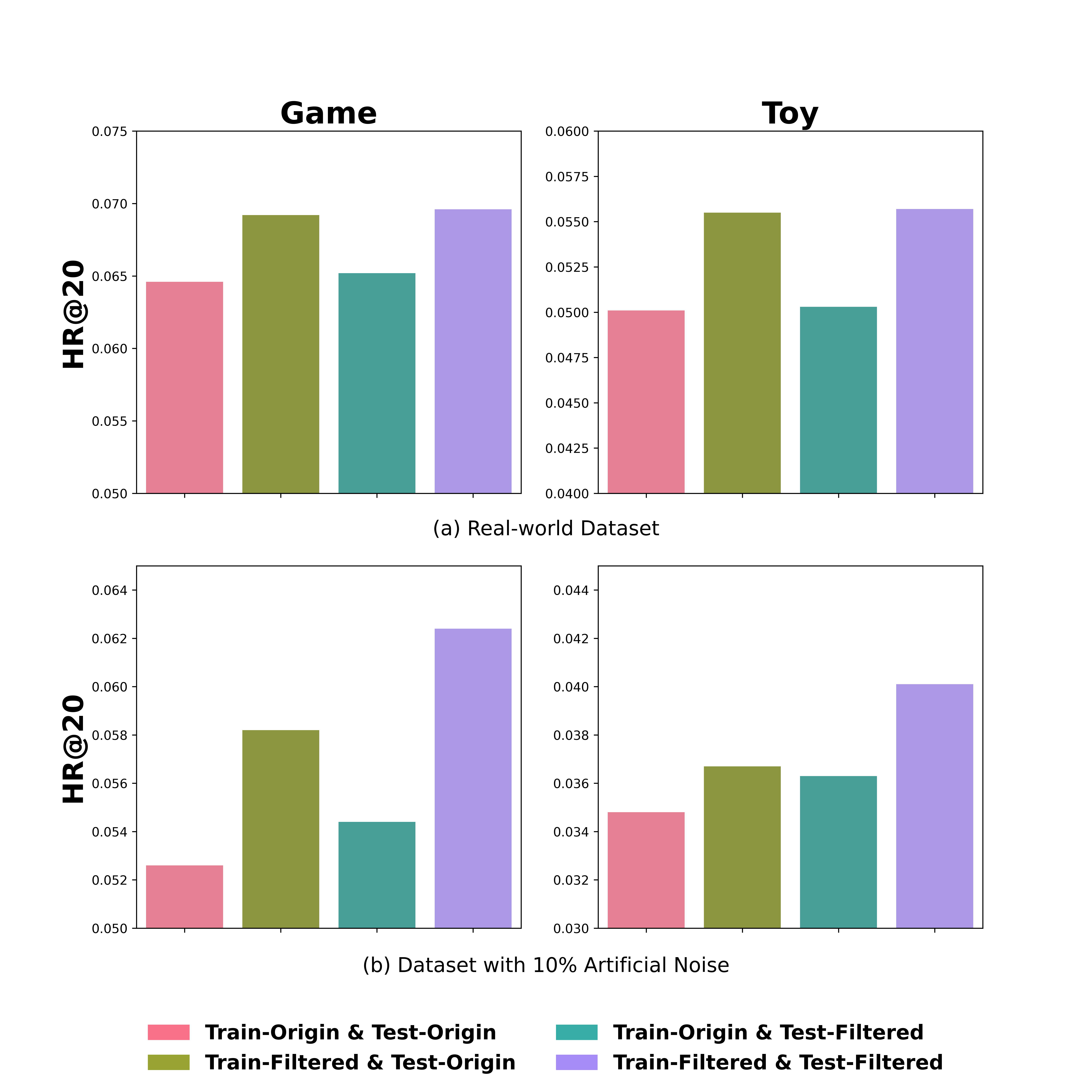}
  \caption{The performance of LLM4DSR for denoising the train and test sets respectively.  Figure (a) shows the experimental results on the natural noise setting, and Figure (b) shows the experimental results on the artificial noise setting with 10\% artificial noise. The experiment is conducted with SASRec as the backbone.}
  \Description{}
  \label{train_test}
\end{figure}

\subsubsection{Performance of separately denoising on training and test sets.}
To more effectively demonstrate the impact of LLM4DSR on model performance, we evaluated the model after denoising the training and testing sets separately. As depicted in Figure \ref{train_test}, we conducted experiments on the Game and Toy datasets. Figure \ref{train_test}(a) presents the results on real-world datasets, while Figure \ref{train_test}(b) shows results on datasets with 10\% artificial noise. Across all datasets and settings, denoising both the training and testing sets individually enhanced performance, with greater improvements observed when denoising the training set. Furthermore, simultaneously denoising both sets led to the most significant performance gains. The results highlight the importance of clean data in both training and evaluation phases, and demonstrate that LLM4DSR can be a valuable tool in enhancing data quality, ultimately leading to better model performance.


\begin{table}[t]
\abovecaptionskip=-0pt
\belowcaptionskip=-10pt
\centering
\caption{The impact of different components in LLM4DSR. "w/ artificial noise" indicates a dataset with artificially added noise and noise ratio is 10\%.}
\label{tab:ablation_study2}
\resizebox{0.6\columnwidth}{!}{%
\begin{tabular}{@{}llcc@{}}
\toprule
Dataset                                     & Method                           & NDCG@20         & HR@20           \\ \midrule
\multirow{5}{*}{Game}                       & SASRec                           & 0.0299          & 0.0646          \\
                                            & w/o uncertainty          & 0.0309          & 0.0648          \\
                                            & w/o completion           & 0.0241          & 0.0566          \\
                                            & w/ popularity completion & 0.0239          & 0.0460          \\
                                            & \textbf{LLM4DSR}                          & \textbf{0.0315} & \textbf{0.0696} \\ \midrule
\multirow{5}{*}{Toy}                        & SASRec                           & 0.0190          & 0.0501          \\
                                            & w/o uncertainty          & 0.0191          & 0.0501          \\
                                            & w/o completion           & 0.0185          & 0.0478          \\
                                            & w/ popularity completion & 0.0156          & 0.0417          \\
                                            & \textbf{LLM4DSR}                          & \textbf{0.0208} & \textbf{0.0557} \\ \midrule
\multirow{5}{*}{\begin{tabular}[c]{@{}l@{}}Game   \\ w/ artificial noise\end{tabular}} & SASRec                           & 0.0252          & 0.0526          \\
                                            & w/o uncertainty          & 0.0277          & 0.0572          \\
                                            & w/o completion           & 0.0264          & 0.0548          \\
                                            & w/ popularity completion & 0.0241          & 0.0472          \\
                                            & \textbf{LLM4DSR}                          & \textbf{0.0292} & \textbf{0.0624} \\ \midrule
\multirow{5}{*}{\begin{tabular}[c]{@{}l@{}}Toy   \\ w/ artificial noise\end{tabular}}  & SASRec                           & 0.0139          & 0.0348          \\
                                            & w/o uncertainty          & 0.0133          & 0.0361          \\
                                            & w/o completion           & 0.0142          & 0.0359          \\
                                            & w/ popularity completion & 0.0106          & 0.0267          \\
                                            & \textbf{LLM4DSR}                          & \textbf{0.0152} & \textbf{0.0401} \\ \bottomrule
\end{tabular}%
}
\end{table}

\begin{figure}[t]
  \centering
  \abovecaptionskip=-0pt
  \belowcaptionskip=-10pt
  \includegraphics[width=\linewidth]{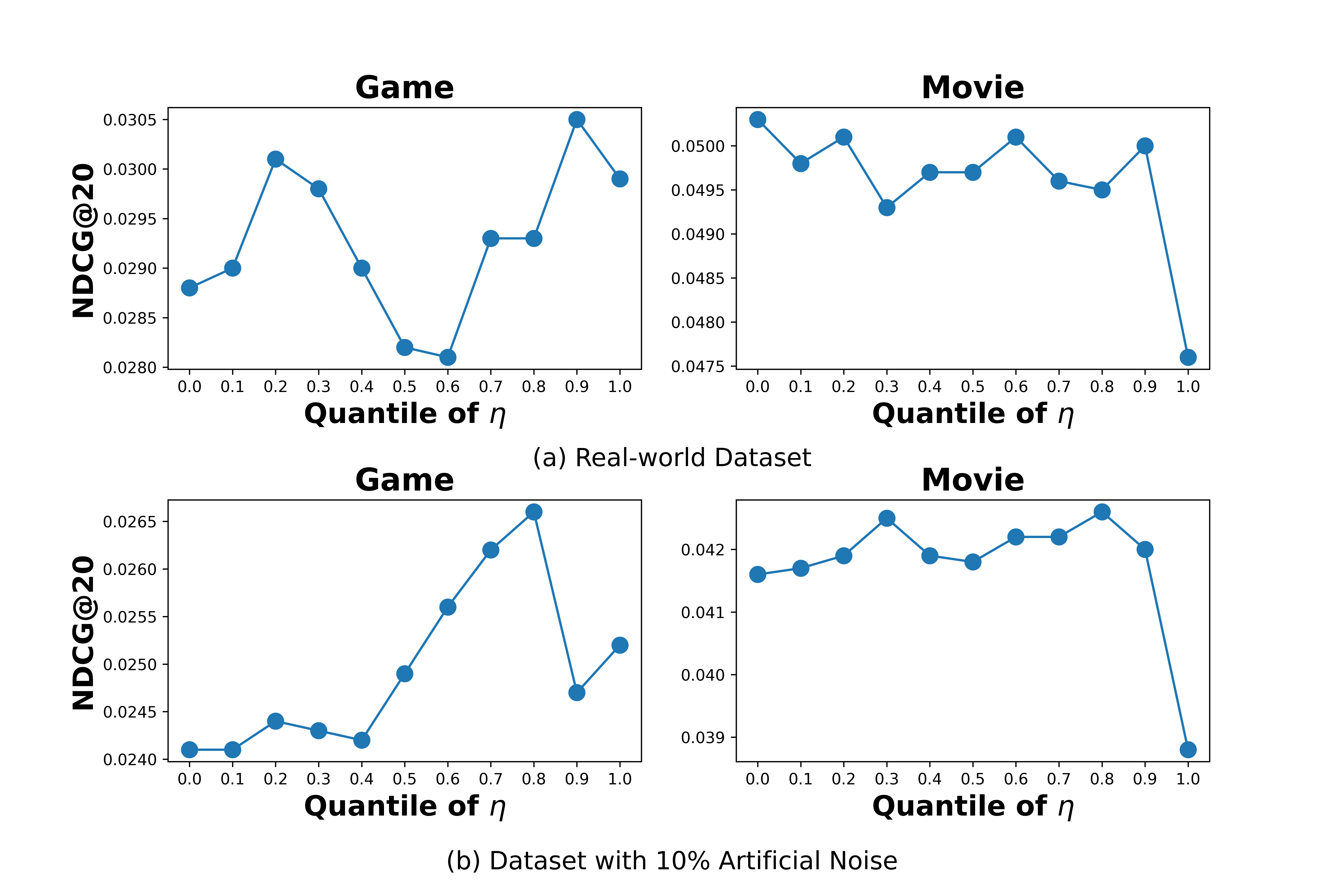}
  \caption{Sensitivity analysis of hyperparameter $\eta$.}
  \Description{}
  \label{train_noise_threshold}
\end{figure}

\subsection{Sensitivity analysis of hyperparameter $\eta$ (RQ3)}
\label{sec:train_noise_threshold}
In the Uncertainty Estimation module, \(\eta\) serves as the threshold distinguishing classification noise from clean samples. As \(\eta\) increases, fewer items are identified as noise, leading to a lower modification ratio of the original sequence. To better understand the impact of \(\eta\) on model performance, we examined the relationship between the hyperparameter \(\eta\) and model performance under two noise settings, as illustrated in Figure \ref{train_noise_threshold}. The x-axis represents the quantile of \(\eta\); a quantile of 1 indicates that all samples are classified as clean, equivalent to training on the original dataset, whereas a quantile of 0 indicates that all samples are identified as noise and replaced (note that we limit the maximum number of modified items per sequence to $b$). We adjusted \(\eta\) only for the training set, without filtering noise from the test data, to accurately reflect the impact of parameter changes. Figure \ref{train_noise_threshold}(a) shows the results on the natural noise setting, while Figure \ref{train_noise_threshold}(b) presents the results on the artificial noise setting.

In the artificially noisy Game dataset, NDCG@20 initially increases and then decreases as \(\eta\) decreases. During the initial increase, LLM4DSR effectively filters out a large number of high-confidence noise samples; subsequently, performance declines due to modifications of some clean samples. In the real-world game dataset, NDCG@20 shows a second drop, which may be due to the fact that a large number of noise samples are concentrated in the samples with medium confidence.
In the Movie, NDCG@20 increases with a decrease in \(\eta\) and then remains stable, suggesting that LLM's replacement operations effectively compensate for the information loss caused by erroneous deletions, thus enhancing the robustness of the \(\eta\) parameter. 
Overall, \(\eta\) is relatively insensitive to adjustments, and satisfactory results can be achieved by searching within a larger \(\eta\) range (e.g., above the 0.7 quantile) during the actual application.

\subsection{Case Study (RQ4)}
\begin{figure}[t]
  \centering
  \includegraphics[width=0.7\linewidth]{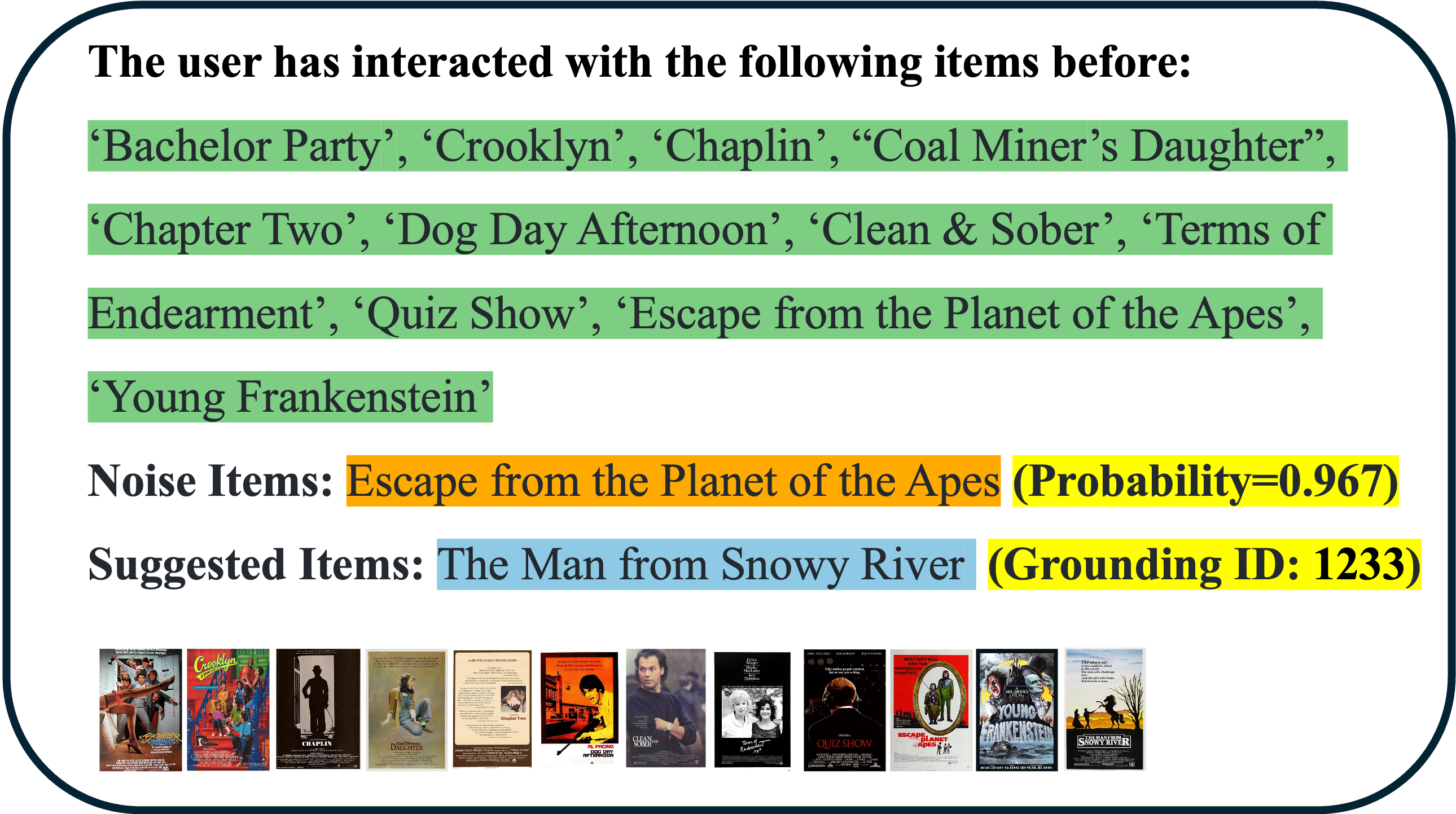}
  \caption{A case study on \textit{Amazon Movies} shows how LLMs use open knowledge to identify and replace noise.}
  \Description{}
  \label{case_study}
\end{figure}

In this case study, we illustrate how LLM4DSR leverages open knowledge to achieve denoising. As depicted in Figure \ref{case_study}, we analyze a user's viewing history from the Movie dataset, which predominantly consists of films in the drama, romance, and comedy genres. A notable outlier in this collection is the science fiction film \textit{"Escape from the Planet of the Apes"}. Utilizing prior knowledge about movie classifications, the LLM identifies this film as noise with a probability of $0.967$. To enhance the sequence information, LLM4DSR recommends \textit{"The Man from Snowy River"}, a film that aligns with the user's preferred drama/romance genres, as a suitable replacement for the identified noise.
Traditional ID-based denoising methods encounter substantial challenges in identifying noise due to their lack of access to semantic and categorical information about items, as well as the absence of supervisory signals from noisy labels. Moreover, these models do not possess reasoning and comprehension capabilities, further complicating the process of noise identification.

\section{Related Work}
\label{Related work}
\subsection{Sequential Recommendation}
The sequential recommendation infers the next item of interest to the user based on their historical interactions. Compared to collaborative filtering, sequential recommendation takes into account the temporal order of interactions to better capture the dynamic changes in user interests. Benefiting from the development of deep learning models in recent years, sequential recommendation employs various deep models to model users' historical interactions and capture their interests. For instance, GRU4Rec \cite{hidasi2015session} uses RNN, while Caser \cite{tang2018personalized} uses CNN. SASRec \cite{kang2018self} and BERT4Rec \cite{sun2019bert4rec} are based on the self-attention mechanism \cite{vaswani2017attention}, which automatically learns the weight of each interaction. 

In addition to innovations in model architecture, recent sequential recommendation models try to address the challenge of capturing rapidly changing user interests over time. For example, DROS \cite{yang2023generic} introduces distributionally robust optimization (DRO) \cite{rahimian2019distributionally} to enhance the model's ability to handle out-of-distribution (OOD) problems caused by temporal changes. SURGE \cite{chang2021sequential} uses GNNs to capture dynamic user interests within implicit interactions.

Furthermore, some studies have incorporated contrastive learning into sequential recommendations to alleviate the issue of sparse recommendation data. Such methods primarily focus on how to construct positive and negative samples within the contrastive loss. CL4SRec \cite{xie2022contrastive} constructs positive samples of the original sequence through sequence transformations such as item cropping, item masking, item reordering. DuoRec \cite{qiu2022contrastive} proposes a model-level data augmentation method based on dropout to achieve better semantic retention. DCRec \cite{yang2023debiased} introduces a contrastive learning loss based on conformity and interest disentanglement to address the popularity bias present in data augmentation. SoftCSR \cite{zhang2024soft} extends contrastive learning to regional-level sample comparisons, offering more flexibility than single-sample contrasts, and employs adversarial contrastive loss to enhance model robustness.

The readers may refer to the excellent survey \cite{fang2020deep, wang2019sequential} for more details. However, the aforementioned methods do not consider the negative impact of noisy interactions on modeling the sequence, which is quite common in real-world scenarios.

\subsection{Denoising Recommendation}

In this section, we first review the current sequential recommendation denoising schemes, followed by a discussion of denoising recommendation under collaborative filtering scenarios.
\subsubsection{Denoising sequential recommendation}
The primary objective of denoising sequential recommendation is to enhance the performance of recommendation systems by mitigating the impact of noisy data. Denoising methods for sequential recommendation can be broadly categorized into implicit denoising and explicit denoising. Implicit denoising methods primarily reduce the impact of noise without explicitly removing them. Some works reduce the influence of noisy interactions to the sequence representation by increasing the sparsity of attention weights in attention-based models \cite{yuan2021dual, chen2022denoising}. FMLP and END4Rec \cite{zhou2022filter, han2024end4rec} consider using filters to remove noise signals from representations. The drawback of these methods is that noise still affect the model, and their high coupling with the model architecture makes it difficult to transfer them to other backbones or downstream tasks, including the widely researched LLM-based recommendation systems in recent years.

Explicit denoising methods directly modify the sequence content, effectively addressing the limitations of implicit denoising techniques. These methods often rely on heuristically designed denoising modules or metrics. BERD \cite{sun2021does} models the uncertainty of each interaction and eliminates modules with high loss but low uncertainty to achieve denoising. HSD \cite{zhang2022hierarchical} assesses the similarity of each sample to other items in the sequence and its alignment with user interests to detect noise. STEAM \cite{lin2023self} trains a discriminator model and a generator model, which are used to identify noise and complete sequences, respectively. 

Regardless of whether explicit or implicit denoising methods are used, the most significant challenge stems from the lack of effective supervision signals from noise labels to train the denoising model. Existing methods often rely on heuristic designs inspired by prior knowledge to develop denoising algorithms, which requires substantial expert knowledge and lacks precision. Additionally, since these methods usually utilize supervision information from recommendation loss, they are still adversely affected by noise as the labels of the recommendation label (i.e., the next item) could also be contaminated. 
Our proposed method, LLM4DSR, addresses these issues by leveraging the extensive open knowledge of Large Language Models (LLMs) to achieve more effective denoising.

\subsubsection{Denoising recommendation under collaborative filtering scenarios.}
Considering other recommendation scenarios, most of work focuses on denoising under collaborative filtering scenarios. Due to the lack of labeled noise data, some works utilize prior knowledge about noise samples to aid in denoising for implicit feedback data. For instance, ADT \cite{wang2021denoising} and SGDL \cite{gao2022self} utilize the phenomenon that models tend to fit clean samples more easily during the early stages of training, while noisy samples exhibit higher loss, to design denoising strategies. ADT simply sets a threshold to reduce the weights of samples with higher loss, whereas SGDL uses the information from the clean samples that the model has fitted to help train a sample weight adjustment model. DeCA \cite{wang2022learning} identifies noise by observing that different models exhibit significant prediction differences for noise samples. However, such prior knowledge is not always reliable. For example, Shu \etal \cite{shu2019meta} points out that samples with high loss are not necessarily noise and could also be hard samples. RGCF \cite{tian2022learning} and GraphDA \cite{fan2023graph} utilize node features extracted by GNNs to reconstruct the graph structure based on feature similarity, aiming to reduce the impact of noise signals in the graph. BOD \cite{wang2023efficient} employs a bi-level optimization approach to train a weight generator that automatically learns the weight of each sample. These methods often involve complex module designs and, due to the lack of labeled noise signals, their performance is still not guaranteed. 


To the best of our knowledge, a recent study has also explored the use of LLMs for recommendation denoising in the collaborative filtering scenario \cite{song2024large}. However, our approach, LLM4DSR, significantly differs from LLMHD in several key aspects: (1) Different tasks: LLM4DSR addresses the task of sequential recommendation, while LLMHD focuses on collaborative filtering. Specifically, our work targets denoising sequential data, whereas LLMHD aims to differentiate between noisy samples and hard negative samples; (2) Different methods: LLM4DSR directly employs instruction tuning and uncertainty estimation to tackle the denoising task, while LLMHD directly uses untuned LLMs and  crafts specific prompts to reassess samples, identifying those that deviate from traditional model predictions as noise. Moreover, LLMHD \cite{song2024large} is a preprint available on arXiv as of September 16, 2024, and has not been formally published. Therefore, we consider our work to be parallel to LLMHD.

\subsection{LLMs for Recommendation}
Large Language Models (LLMs), with their extensive knowledge, generalization capabilities, and reasoning abilities, have achieved remarkable results across numerous fields. 
LLMs are utilized in various recommendation tasks, including collaborative filtering \cite{zhu2024collaborative, wang2024enhanced}, sequential recommendation \cite{hou2023learning, hou2022towards, bao2023bi}, graph-based recommendation \cite{wei2024llmrec}, and CTR (Click-Through Rate) tasks \cite{bao2023tallrec, wang2023alt, xu2024enhancing}. Additionally, LLMs have been utilized for tasks that were challenging for traditional model to achieve, such as explaining item embeddings \cite{tennenholtz2023demystifying}, explaining recommendation results \cite{gao2023chat, wang2024can}, and conversational recommendations \cite{gao2023chat}.

The initial attempts directly use pre-trained LLMs as the backbone for recommendations to explore their zero-shot capabilities \cite{gao2023chat, hou2024large, liu2024once, wang2023zero, wang2024recommend}. These methods perform not well due to the significant gap between recommendation tasks and the training tasks of LLMs. Subsequent research has organized recommendation data into prompt formats and then fine-tuned LLMs to enhance their recommendation performance, achieving better results \cite{bao2023bi, bao2023tallrec, hu2024exact, li2023prompt, li2024citation, li2023exploring, li2024pap, shi2024enhancing, wang2024towards, zhang2023recommendation, zhang2024tired, gao2024end}. 

Besides generating recommendation results directly, LLMs can also serve as enhancers to improve the performance of conventional recommenders. Existing methods mainly employ LLMs as encoders to encode the semantic information of users and items into embeddings \cite{xi2023towards, ren2024representation, wei2024llmrec, wang2023alt} or as an additional knowledge base \cite{yang2024common, qin2024d2k}. However, such methods largely fail to leverage the inherent reasoning capabilities of LLMs. To address this shortcoming, SLIM \cite{wang2024can} has enabled LLMs to generate Chain-of-Thought (CoT) data \cite{wei2022chain} as additional input for downstream models. Since the content generated by LLMs is not optimized with recommendation task, this kind of information may not necessarily benefit traditional models. To address this, SeRALM \cite{ren2024enhancing} considered optimizing the outputs of LLMs using recommendation loss by alignment training.


To the best of our knowledge, this paper is the first to explore the application of LLMs in the field of denoising sequential recommendation.

\section{Conclusion}
\label{Conclusion}
In this paper, we propose LLM4DSR, a novel method for sequence recommendation denoising using Large Language Models (LLMs), which is the first exploration of LLM performance on this task. Leveraging the extensive open knowledge embedded in LLMs, LLM4DSR offers significant advantages over traditional denoising methods. We have designed a self-supervised task that endows LLMs with enhanced denoising capabilities and integrated the Uncertainty Estimation module to improve both the accuracy and flexibility of the denoising process. Furthermore, we utilized the generative capabilities of LLMs to complete the denoised sequences, thereby enriching sequence information and ensuring robust performance in downstream tasks. Comprehensive experiments demonstrate that LLM4DSR outperforms existing state-of-the-art sequence denoising techniques.

Despite the notable performance of LLM4DSR, it incurs higher computational costs compared to traditional methods due to the use of LLMs. A promising direction for future research is to employ techniques such as knowledge distillation to transfer the denoising capabilities of LLMs to smaller models, thereby achieving superior performance without increasing computational overhead.

\bibliographystyle{ACM-Reference-Format}
\bibliography{sample-base}

\end{document}